\documentclass{ieeeaccess}

\usepackage{cite}
\usepackage{algorithmic}
\usepackage{graphicx}
\usepackage{textcomp}
\usepackage{hyperref}
\usepackage{amsfonts,amsmath,amsthm,amssymb,mathtools,mathrsfs}
\usepackage{url,array}
\usepackage{threeparttable, booktabs}
\usepackage{xcolor}

\usepackage{tabularx} 

\renewcommand{\Re}{\operatorname{Re}}
\newcommand{\pos}{\mathrm{pos}}
\renewcommand{\neg}{\mathrm{neg}}

\newcommand{\rline}  {{\mathbb R}}
\renewcommand{\cline}{{\mathbb C}}

\newcommand{\dd}  {{\rm d}\hbox{\hskip 0.5pt}}
\renewcommand{\o} {{\omega}}
\renewcommand{\l} {{\lambda}}
\newcommand{\half}{{\frac{1}{2}}}
\renewcommand{\leq} {\leqslant}
\renewcommand{\geq} {\geqslant}

\newcommand{\Hscr} {{\cal H}}

\newcommand{\e}    {{\varepsilon}}

\newcommand{\m}    {{\hbox{\hskip 1pt}}}

\newcommand{\nm}   {{\hbox{\hskip -3pt}}}
\newcommand{\bbm}[1]{\left[\begin{matrix} #1 
            \end{matrix}\right]}
\newcommand{\sbm}[1]{\left[\begin{smallmatrix} #1 
            \end{smallmatrix}\right]}

\newtheorem{theorem}{Theorem}

\newtheorem{proposition}[theorem]{Proposition}

\usepackage{bm}
\makeatletter
\AtBeginDocument{\DeclareMathVersion{bold}
\SetSymbolFont{operators}{bold}{T1}{times}{b}{n}
\SetSymbolFont{NewLetters}{bold}{T1}{times}{b}{it}
\SetMathAlphabet{\mathrm}{bold}{T1}{times}{b}{n}
\SetMathAlphabet{\mathit}{bold}{T1}{times}{b}{it}
\SetMathAlphabet{\mathbf}{bold}{T1}{times}{b}{n}
\SetMathAlphabet{\mathtt}{bold}{OT1}{pcr}{b}{n}
\SetSymbolFont{symbols}{bold}{OMS}{cmsy}{b}{n}
\renewcommand\boldmath{\@nomath\boldmath\mathversion{bold}}}
\makeatother

\def\BibTeX{{\rm B\kern-.05em{\sc i\kern-.025em b}\kern-.08em
    T\kern-.1667em\lower.7ex\hbox{E}\kern-.125emX}}

\begin{document}

\history{Date of publication xxxx 00, 0000, date of current version 
   xxxx 00, 0000.}
\doi{10.1109/ACCESS.2024.0429000}

\title{Generalized $H$ matrix stability for power systems}
\author{\uppercase{Elinor Ginzburg-Ganz}\authorrefmark{1}, 
   \uppercase{Yoash Levron}\authorrefmark{1}, 
   \IEEEmembership{Senior Member, IEEE},
   \uppercase{juri Belikov}\authorrefmark{2}, 
   \IEEEmembership{Senior Member, IEEE}, and
   \uppercase{George Weiss}\authorrefmark{3}, 
   \IEEEmembership{Member, IEEE}}

\address[1]{The Andrew and Erna Viterbi Faculty of Electrical and 
   Computer Engineering, Technion, Israel Institute of Technology, 
   Haifa 3200003, Israel (e-mail: yoashl@ee.technion.ac.il)}
\address[2]{Department of Software Science, Tallinn University of 
   Technology, Akadeemia tee 15a, 12618 Tallinn, Estonia 
   (juri.belikov@taltech.ee)}
\address[3]{The School of  Electrical and Computer Engineering, Tel-Aviv
   University, Ramat Aviv 69978, Israel (gweiss@tauex.tau.ac.il)}
\tfootnote{The work was partly supported by the Israel Science 
   Foundation and the Estonian Research Council (grant no. 
   PRG1463).\\
   The last author’s research has been supported by a grant from Mr
   Jack Trabin (USA).}

\markboth
{Author \headeretal: Preparation of Papers for IEEE Access}
{Author \headeretal: Preparation of Papers for IEEE Access}

\corresp{Corresponding author: Yoash Levron 
        (e-mail: yoashl@ee.technion.ac.il).}

\begin{abstract}
The small-signal stability of a power system model with arbitrary
loads is investigated, with a focus on locally constant power loads in
lossless networks. A central element of the proposed analysis is the
matrix $H$, which represents the partial derivatives of generator
powers with respect to their relative angles. The local stability of
any equilibrium point is determined by the properties of $H$ if the
loads consume locally constant power. Specifically, an equilibrium
point is asymptotically stable if $H$ is positive definite and
unstable if $H$ is full-rank and not positive definite, while several
quantitative relationships between the eigenvalues of $H$ and the
system Jacobian matrix are also established. The main contribution is
that loads are allowed to be arbitrarily located within the
transmission network, thereby generalizing the existing $H$ matrix
stability criterion which is valid only if loads are located at the
same buses as the generators. The advantage of the proposed criterion
is demonstrated through case studies involving systems with up to 2869
buses. Accurate stability assessments are maintained, where
alternative methods, such as the NRPS model, are found to be overly
conservative. Additionally, a computational acceleration by a factor
of 340 relative to Jacobian-based methods is achieved.
\end{abstract}

\begin{keywords}
$H$ matrix, multi-machine power system dynamics, NRPS model, positive 
matrix, power system stability, small signal stability, time-varying 
phasors.
\end{keywords}

\titlepgskip=-21pt

\maketitle

\section*{Notations} 
\begin{tabularx}{\columnwidth}{@{} l @{\ --\ } X}
$M$ & matrix of generators' inertia constants \\
$D$ & matrix of generators' damping coefficients \\
$U$ & matrix mapping generator angles to relative rotor angles \\
$H$ & matrix of partial derivatives of generator active powers w.r.t. 
      relative rotor angles \\
$\mathcal{H}$ & full Jacobian matrix of active powers w.r.t. relative 
      rotor angles \\
$Y$ & admittance matrix of the transmission network at the nominal 
      frequency\\
$A$ & system matrix of the linearized dynamic system \\
$\omega$ & vector of generator electric frequencies relative to the 
      reference frequency \\
$V_i$ & voltage amplitude at bus $i$\\
$P_i$ & active power injected at bus $i$ \\
$X_-(A)$ & strictly stable subspace of matrix $A$ \\
$X_+(A)$ & strictly unstable subspace of matrix $A$ \\
$\sigma(A)$ & the set of eigenvalues of matrix $A$ \\
\end{tabularx}

\section{Introduction} 

Within the field of electrical engineering, power system stability
remains a primary focus of contemporary research, primarily driven by
the increasing penetration of distributed energy resources (DERs) and
the ongoing structural evolution of modern electrical grids. These
stability challenges are exacerbated by the inherent complexity and
heterogeneity of the underlying power system models
\cite{MilanDHHV2018,Stegink:19}.

The stability of power system models has been researched for many
years, using many methods and points of view, see for instance
\cite{JePeSc:23,BelikL2018,ReYiWe:22,BSOE:17,refylquasidq0,
WeissDL2019,CGBD:18,Schiff:19}. The literature in this field is
extensive (the aforementioned citations representing only a subset)
and originates from diverse research disciplines over many years. In
general, finding sufficient conditions for the stability of a power
network is extremely difficult. This is because power networks are
nonlinear and high-dimensional systems. Every type of analysis is
based on an approximate model of the system. For instance, an
individual synchronous generator can be modeled with 0, 1, or 2 damper
windings, with constant field current or with varying field current,
and so forth. The complexity of synchronous machine models ranges from
order 9, see for instance \cite[Chapter~4]{AnderF2008} or the survey
\cite{Stegink:19}, down to order $2$, see for instance
\cite{DoBu:12,Schiff:19}. Likewise, prime movers and their governors,
as well as loads, can be modeled with varying complexity \cite{WuLiu,
GrArDo:16,CGBD:18,DoBu:12,Schiff:19,art167}. Clearly, the more complex
the model of the power network, the more difficult it is to prove
stability results for it. While computational methods may be used to
simulate the behaviour of any specific system \cite{TLV-KT:15}, such
methods are generally non-scalable, and rarely lead to deep insights.

The approximation of time-varying phasors is commonly used to model
dynamic phenomena in large-scale power systems \cite{GrainS1994}. This
approach describes voltages and currents with time-varying phasor
amplitudes and phases, allowing the transmission network to be modeled
using power-flow equations and simplified phasor-based models for
generators and loads. The core assumption is that system transients
are significantly slower than the nominal system frequency
\cite{IlicZ2000,StefaS2002}. While this generally yields good results
for slow transients in large systems, its accuracy may diminish with
high renewable energy penetration or in systems with fast transients
\cite{IlicZ2000,IlicJM2017,DemirA2006a,BelikL2018b,LeNw:09,
UlbigBA2014}. Still, time-varying phasor models are
very popular for several reasons. Firstly, they use intuitive
quantities like amplitude, phase, active power, and reactive power,
which power engineers readily understand as they stem directly from
phasor notations. Secondly, these models are simpler and more compact
numerically than other, more complex models, making them
computationally efficient. Thirdly, time-varying phasor models are
remarkably flexible, capable of describing a wide variety of sources
and loads, with numerous models already documented in
literature. Finally, their main assumption of slow transients holds
true for most large-scale network explorations.

Within the framework of time-varying phasors, the dynamic stability of
power systems has also been studied extensively. A few examples
include the use of Kuramoto models \cite{DoBu:12,DoBu:14,WeissDL2019,
FD-FB:11,STROGATZ20001}, network-reduced power system models (NRPS)
\cite{ReYiWe:22, art167}, DC power flow \cite{LiTrWe:20,JePeSc:23},
linearization techniques \cite{AnderF2008}, and Lyapunov functions
\cite{SiAlBr:03,TLV-KT:15,chu2005constructing}, among
others. Alternatively, predictive methods have been proposed to assess
small-signal stability, particularly when system configurations are
complex or unknown. For instance, \cite{Jiang2026An} introduces an
automated dynamic frequency scanning tool designed for the analysis of
systems with unknown internal configurations. By injecting multi-sine
disturbances into an electromagnetic transient environment and
analyzing the resulting waveforms, the tool extracts the small-signal
frequency-dependent impedance of the system at a specific operating
point. This enables the characterization of the small-signal frequency
response and the prediction of stability margins without requiring
access to the underlying control structures or internal parameters of
the components. These models serve as practical tools for real-world
analysis, as demonstrated in \cite{Lin2024Transient}. Gaining a better
understanding of the fundamental characteristics of the matrices used
for stability analysis may be highly relevant for recent research
focusing on active control methods, such as
\cite{Kumar2022Performance, KUMAR2021Multimachine} and
\cite{Xygkis2025Investigation}. The properties of the $H$ matrix may
establish a practical direction for such control strategies, where the
synchronizing power coefficients are used as a real-time stability
objective. Within this framework, virtual inertia and damping
parameters might be adaptively tuned to ensure that the $H$ matrix
remains positive definite. One prominent method for assessing the
small-signal stability of power systems that use phasor-based dynamic
models is the $H$ matrix stability criterion, as detailed for instance
in \cite{MachoBB2011} and \cite{Li2024Small}. The classical
application of this criterion is specifically designed for a
particular power system model, that incorporates time-varying phasors,
uses second order models for generators, assumes a lossless
transmission network, and considers constant power loads. In this
context, the matrix $H$ is computed at a certain equilibrium point
$x^*$, and is constructed from the partial derivatives of generator
powers with respect to their relative angles. The stability of the
equilibrium point is then determined as follows: If $H$ is positive
definite, the equilibrium point is asymptotically stable, and if $H$
is full-rank but not positive definite, the equilibrium point is
unstable (see \cite{MachoBB2011}).

Recently, data-driven frameworks such as the one presented in
\cite{Shuai2025Physics}, are proposed to learn power system dynamics
by incorporating physical laws into the network architecture. In
comparison to such data-driven approaches, the $H$ matrix stability
criterion provides analytical rigor through a necessary and sufficient
condition for stability (under the assumptions stated in Section
\ref{sec:Problem}), where data-driven frameworks serve as a predictive
tool, whose accuracy is dependent on training data, and can suffer
from loss of previously learned dynamics when transitioning between
stable and unstable regimes. The $H$ matrix method also exhibits
higher online computational speed, as the assessment is performed
directly without the requirement for intensive training. For example,
when examining a large-scale system, such as the 2869-bus system, the
stability of the equilibrium point is determined in 0.0011 seconds
using the $H$ matrix stability criterion. Alternatively, the proposed
data-driven method, while more efficient than traditional data-driven
approaches, still requires 0.024 seconds to solve a small 4-bus
system. Furthermore, the $H$ matrix stability criterion is much more
interpretable than any data-driven methods.

While a detailed comparison of the proposed stability criterion to
previous results may be found in Section \ref{sec:compare}, the most
notable advancement lies in the topology of the transmission
network. Previous works, for instance \cite{MachoBB2011}, assume that
loads are situated on the same buses as generators. The proposed
updated criterion eliminates this assumption, allowing loads to be
located anywhere in the transmission network. This enhancement enables
the $H$ matrix criterion to be applied to any lossless transmission
network model, making it more general and broadly applicable than
previously available methods. The rather nontrivial fact that under
the stated assumptions $H$ is symmetric and self-adjoint is also
proven. Furthermore, several other relations between the eigenvalues
of $H$ and the eigenvalues of the system's Jacobian matrix $A$ are
established. For instance, it is shown that, under certain conditions,
the number of unstable eigenvalues of $A$ equals the number of
negative eigenvalues of $H$ when algebraic multiplicities are
counted. A number of case studies, both theoretical and numerical,
that demonstrate the usefulness of the proposed updated criterion are
also explored.

{\bf The main result}, Theorem~\ref{main}, is an updated and
generalized $H$ matrix stability criterion for phasor-based dynamic
power grid models with locally constant power loads. Various
extensions to arbitrary power loads are also derived in Section
\ref{sec:generalize}. In the model, each generator is represented by a
swing equation, as in \eqref{eq:dynamic_system1}.  Such a generator
may also be an inverter controlled to emulate a synchronous generator,
which is an often employed control strategy for grid forming
inverters, see \cite{ZhWe:11}, \cite{art150}, \cite{art167}.

The contributions of the paper are summarized as follows:
\begin{enumerate}
    \item An updated stability criterion is developed for the analysis
    of small-signal stability in power systems with arbitrary load 
    distributions.
    \item The symmetry of the $H$ matrix is formally established for
    lossless transmission networks with locally constant power loads.
    \item Quantitative relationships between the eigenvalues of the 
    $H$ matrix and the system Jacobian are derived to characterize 
    stability and instability.
\end{enumerate}
The advantage of the proposed criterion is demonstrated by several
case studies, which show that correct stability assessments are
maintained in large-scale power systems where alternative
methodologies, such as network reduction methods, are found to be
overly conservative and do not yield unequivocal results. This
limitation is exemplified in the 2869-bus Pegase system, where the
network reduction approaches established in \cite{DoBu:12} and
\cite{WeissDL2019} are unable to determine whether the system is
stable. From a different perspective, although correct results are
achieved by the methodology proposed in \cite{WuLiu}, through the use
of the full Jacobian matrix, the process is considerably
slower. Specifically, the $H$ matrix calculation is performed over 340
times faster than the Jacobian-based approach for this system. These
efficiency gains are obtained because the $H$ matrix is characterized
by a dimension approximately half that of the system matrix $A$, which
results in substantial reductions in both computational time and
memory requirements.

The paper continues as follows: In Section~\ref{sec:Problem}, basic
definitions and the problem formulation are provided. It is also
proven that in the case of locally constant power loads, $H$ is
symmetric. Section~\ref{sec:main_result} contains the main result,
Section~\ref{sec:generalize} generalizes the result for arbitrary
loads, and Section~\ref{sec:compare} discusses the relation between
the main result of this paper and related earlier results.
Sections~\ref{sec:case_study_typeI}--\ref{sec:case_study_numerical}
provide case studies.

A short conference version of this material (with some 
proofs omitted and with different examples) is in \cite{art168p5}.

\section{Problem formulation} \label{sec:Problem} 

\begin{figure}[t!] \centering 
   \includegraphics[width=0.8\columnwidth]{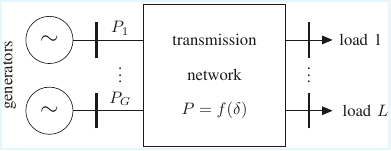}
   \caption{A general power system model, presenting $G$ generators 
   and $L$ loads. The lossless transmission network is described by 
   the algebraic function $P=f(\delta)$ that links relative generator 
   angles to their active power output. A similar model is used in 
   \cite[p.~36]{AnderF2008}, \cite[p.~514]{MachoBB2011}.} 
   \label{fig:gen_net}
\end{figure}

A general balanced AC power system is considered, whose dynamic
behavior is governed by the quasi-stationary model, as considered for
instance in \cite{chu2005constructing,GrainS1994,MachoBB2011}. Note
that this paper relies only on the quasi-static model of a
transmission network with loads. In this model, the consumed active
and reactive powers are functions of the generator voltages and phase
differences only, and there are no differential equations in the
description of the transmission network. The system is illustrated in
Fig.~\ref{fig:gen_net}, and includes a general lossless transmission
network, $G$ generators, $L$ loads and hence $N=G+L$ relevant
buses. Without loss of generality, it is assumed that the generator
buses are labeled by $1,\ldots G$, and the load buses are labeled by
${(G+1)},\ldots N$.

Several assumptions are employed that are commonly used in
time-varying phasor models:
\begin{itemize}
\item The power system is a balanced three-phase system. Transients in
the system are much slower than a 50/60 Hz cycle, and dynamic
phenomena are represented by phasors, based on the quasi-static (or
quasi-stationary) approximation \cite{SchifZOSSR2016}.

\item The transmission network is represented by a linear single-phase
equivalent circuit, and can be described well enough by a constant and
symmetric admittance matrix $Y$, which is calculated at the nominal
system frequency. It is also assumed that the transmission network is
lossless, such that the total generated active power equals the total
active power supplied to the loads. Equivalently, the entries of $Y$
are purely imaginary\footnote{This ensures the symmetry of the $H$
matrix. This assumption is consistent with established models, such as
the one presented in \cite{MachoBB2011}.}.

\item Each generator is represented by the simplest quasi-static
model: a voltage source with constant amplitude behind a series
impedance \cite{GrainS1994}. The generators' synchronous impedances
are modeled as part of the transmission network. The mechanical power
driving the rotor is governed by a linear droop-control mechanism,
where the droop power is a decreasing affine function of the rotor
frequency, with no additional dynamics\footnote{This 
is only an approximation. In practice, the droop control mechanism is
not linear, and may have dead-band, saturation and its own 
dynamics.}.

\item Each load can be decomposed into a power sink, which is
connected in parallel to additional reactive components (inductors and
capacitors). The main assumption for each power sink is that the
consumed active and reactive powers ($P_\ell$ and $Q_\ell$) are
continuously differentiable functions of only the local load's voltage
amplitude ($V_\ell$), and \vspace{-2mm}
\begin{equation} \label{eq:loadsAA}
   \frac{\text{d}P_\ell}{\text{d}V_\ell} \m=\m 0,\qquad
   \frac{\text{d}Q_\ell}{\text{d}V_\ell} \m=\m 0
\end{equation}
at an equilibrium point of interest $x^*$. This is the case, for
instance, if each load consumes constant active and reactive powers.
The additional reactive components can be regarded as part of the
lossless transmission network.
\end{itemize}

The convention that all the voltage phasor angles $\delta_2, \delta_3, \ldots, \delta_N$ are measured with respect to the first generator, so that $\delta_1=0$, is used. Also, angles that differ by an integer multiple of $2\pi$ are identified. Then, the 
following vectors are defined:
\begin{align*}
  \o(t) \m&=\m \left[\o_1(t)-\o_s, \ldots\,\o_G(t)-\o_s 
        \right]^\top, \\
  \delta(t) \m&=\m \left[\delta_2(t), \ldots\,\delta_G(t)
        \right]^\top,
\end{align*}
\begin{align*}
  P(t) \m&=\m \left[P_1(t),\ldots\, P_G(t)\right]^\top, \\
  x(t) \m&=\m \left[\o^\top(t),\, \delta^\top(t)\right]^\top.
\end{align*}
Here, $\o_i(t)$ is the electric frequency of generator $i$ in rad/s,
$P_i(t)$ is the generator's active power output, $\delta_i(t)$ is the
relative rotor angle as defined above, $\o_s$ is the reference
frequency, and $x(t)$ is the state vector. The system is described by
the following differential and algebraic equations \cite{GrainS1994}:
\begin{equation} \label{eq:dynamic_system1}
  \begin{cases}
  M\dot \o(t) + D\o(t) \m=\m {P_{ref}} - P(t), \\ 
  \dot \delta (t) \m=\m U\o(t), \\ P(t) \m=\m f(\delta(t)).
  \end{cases}
\end{equation}

The matrix $M$ represents the generators' inertia constants, and the
matrix $D$ represents the generators' damping coefficients, for
instance due to the damper windings, combined with the prime movers'
droop control constants. Both matrices are symmetric and positive
definite\footnote{In typical time-varying phasor models, the matrices
$D$ and $M$ are also diagonal, but in this paper there is no need for this
property, so $D$ and $M$ are only assumed to be positive definite.
There exists advanced models in which these matrices are not diagonal.
For instance, virtual friction applied to virtual synchronous
machines, as in \cite{ReYiWe:22}, leads to a non-diagonal matrix
$D$.}. The vector $P_{ref}$ contains the reference powers for the
generators. The matrix $U$ describes the relations
$\dot\delta_i=\o_i-\o_1$ for $i=2,\ldots G$, and is given by
\vspace{-2mm}
\begin{equation*}
   U \m=\m \begin{bmatrix} -1_{(G-1)\times 1} & I_{(G-1)\times(G-1)}
   \end{bmatrix},
\end{equation*}
where $1_{(G-1)\times 1}$ is a vector of ones, and $I_{(G-1)\times
(G-1)}$ is the identity matrix. The function $f(\cdot)$ depends on the
transmission network and the loads. This function is usually not
available in closed form, and has to be evaluated numerically, by
solving the network's AC power-flow equations. At an equilibrium point
$x^{*}$ (if one exists), define
\begin{equation} \label{eq:matrixH}
   H \m= \begin{bmatrix} h_{2,2} & \cdots & h_{2,G} \\ \vdots & \ddots
   & \vdots \\ h_{G,2} & \cdots & h_{G,G} \end{bmatrix} =\m 
   \begin{bmatrix} \frac{\partial P_2}{\partial \delta_2} & \cdots & 
   \frac{\partial P_2}{\partial \delta_G} \\ \vdots & \ddots & \vdots 
   \\ \frac{\partial P_G}{\partial \delta_2} & \cdots & \frac{\partial 
   P_G}{\partial \delta_G} \end{bmatrix}.
\end{equation}
Since the transmission network is lossless, it also holds that
\begin{equation} \label{eq:losslessAss}
   \begin{aligned} h_{1,k} &\triangleq \left.\frac{\partial P_1}
   {\partial\delta_k} \right\vert_{x=x^*} = \\ 
   &=\left.\frac{\partial}{\partial\delta_k}\left(P_L - 
   \sum_{i=2}^GP_i \right)\right\vert_{x=x^*} = 
   - \sum_{i=2}^Gh_{i,k}, \end{aligned}
\end{equation}
where $P_L$ is the loads' total active power. Due to \eqref{eq:loadsAA}, it follows that for all $k\in\{2,\ldots G\}$,
\begin{equation*}
   \frac{{\partial {P_L}}}{{\partial {\delta _k}}} \m=\m \sum
   \limits_{\ell=G+1}^N {\frac{{\partial {P_\ell }}} {{\partial
   {\delta_k}}}} \m=\m \sum\limits_{\ell=G+1}^N {{\frac{{\dd
   {P_\ell }}}{{\dd{V_\ell}}} \frac{{\partial {V_\ell }}}
   {{\partial {\delta_k}}}}} \m=\m 0 \m.
\end{equation*} 
From \eqref{eq:losslessAss} it follows that $\partial P/\partial 
\delta = U^\top H$. Based on the definitions above, the Jacobian 
matrix of the dynamic system $\dot{x}=F(x)$ from 
\eqref{eq:dynamic_system1} is
\begin{equation} \label{eq:JacobianA}
   A \m=\m \frac{\partial F}{\partial x} \m=\m \left[ 
   \begin{array}{c|c} -M^{-1}D & -M^{-1} U^{\mathsf{T}} H \\ \hline U 
   & 0_{(G-1)\times (G-1)} \end{array} \right].
\end{equation}

Before stating the main result, it is shown that, under the assumptions
listed above, the matrix $H$ is symmetric:

\begin{theorem} \label{th:symm}
Let $H$ be the matrix in \eqref{eq:matrixH}. If
\begin{itemize}
    \item the admittance matrix $Y$ is purely imaginary,
    \item the loads satisfy \eqref{eq:loadsAA},
    \item the generators' voltage amplitudes are constant,
\end{itemize}
then $H$ is symmetric at $x^*$.
\end{theorem}

{\it Proof.} \m Define $P_i$, $Q_i$ to be the active and reactive
powers injected into bus $i$, and $V_i$, $\delta_i$ to be the voltage
amplitude and phase at bus $i$. In addition, denote the components of
$Y$ by $y_{i,n}$. These variables are linked by the standard AC power
flow equations (\nm\cite{GrainS1994,MachoBB2011,AnderF2008}):
\begin{equation*}
   \begin{aligned} P_i &= \sum_{n=1}^{N}V_i V_n |y_{i,n}|\cos\left(
   \delta_n-\delta_i+\angle y_{i,n}\right), \\
   Q_i &= -\sum_{n=1}^{N}V_i V_n |y_{i,n}|\sin\left(\delta_n-
   \delta_i+\angle y_{i,n}\right), \end{aligned}
\end{equation*}
for $i=1,\ldots N$. Define $J$ to be the Jacobian of these equations
at the equilibrium point $x^*$ (see for instance 
\cite{GrainS1994,MachoBB2011}), such that
\begin{multline*}
   \left[\dd P_2,\ldots \dd P_N,\dd Q_2,\ldots \dd Q_N \right]^\top 
   \m= \\ =\m J\cdot\left[\dd\delta_2,\ldots \dd\delta_N ,\frac{\dd 
   V_2}{V_2^*},\ldots \frac{\dd V_N}{V_N^*} \right]^\top,
\end{multline*}
where $\dd P_i$, $\dd Q_i$, $\dd\delta_i$, $\dd V_i$ are the 
differentials at $x^*$, and $V_i^*$ are the voltage amplitudes at 
$x^*$, for $i=2,\ldots  N$. Note that $\dd V_1$ does not appear here,
since it is assumed that $\dd V_1=0$. Now reorder the rows and 
columns of $J$ to obtain
\begin{equation} \label{eq:JJJ}
   \tilde{J} \begin{bmatrix} r_a \\ r_b \end{bmatrix} \m=\m
   \begin{bmatrix} w_a \\ w_b \end{bmatrix}
\end{equation}
with
\begin{equation*}
   \begin{aligned} r_a^\top &=\m \left[\dd\delta_2,\ldots \dd
   \delta_G,\frac{\dd V_2}{V_2^*},\ldots \frac{\dd V_G}{V_G^*} 
   \right]^\top,\\ r_b^\top &=\m \left[\dd\delta_{(G+1)},\ldots \dd
   \delta_N,\frac{\dd V_{(G+1)}}{V_{(G+1)}^*},\ldots \frac{\dd V_N}
   {V_N^*} \right]^\top,\\ w_a^\top &=\m \left[\dd P_2,\ldots \dd 
   P_G,\dd Q_2,\ldots \dd Q_G \right]^\top,\\ w_b^\top &=\m \left[
   \dd P_{(G+1)},\ldots \dd P_N,\dd Q_{(G+1)}, \ldots \dd Q_N 
   \right]^\top. \end{aligned}    
\end{equation*}
By straightforward calculation, it can be verified that if the
admittance matrix $Y$ is purely imaginary, then the matrix $J$ is
symmetric (see \cite[Chapt.~9]{GrainS1994}, or \cite[p.~117]
{MachoBB2011}). The matrix $\tilde{J}$ may be obtained from $J$ via 
a permutation matrix $S$, such that $\tilde{J}=S J S^\top$, so
$\tilde{J}$ is also symmetric. It follows that \eqref{eq:JJJ} can 
be rewritten as
\begin{equation} \label{eq:almo}
   \begin{bmatrix} \tilde{J}_{1,1} & \tilde{J}_{1,2} \\
   \tilde{J}_{1,2}^\top & \tilde{J}_{2,2} \end{bmatrix}
   \begin{bmatrix} r_a \\ r_b \end{bmatrix} \m=\m \begin{bmatrix}
   w_a \\ w_b \end{bmatrix},
\end{equation}
where $\tilde{J}_{1,1}$ and $\tilde{J}_{2,2}$ are symmetric.  It is
well-known that $\tilde{J}_{2,2}$, being real and symmetric, can be
factored like $\tilde J_{2,2} \m=\m {W^{\mathsf{T}}}\tilde \Lambda W$, 
where $W$ is real, $W^\mathsf{T} W=WW^\mathsf{T}=I$ (unitary matrix),
and $\tilde \Lambda$ is diagonal, with the eigenvalues of $\tilde
J_{2,2}$ (which are real) on its diagonal. The order of the
eigenvalues in $\tilde\Lambda$ is arranged such that the zero eigenvalues (if
they exist) are the last. Thus,
\begin{equation*}
  \tilde \Lambda \m=\m \left[ {\begin{array}{*{20}{c}}
  \Lambda & \vline & 0 \\ \hline 0 & \vline & 0 \end{array}} \right],
\end{equation*}
where $\Lambda$ is a real diagonal matrix of dimension $h\times h$,
$h \in \left\{ {0,1, \ldots ,N - G} \right\}$, with no zeros on its
main diagonal. Equation \eqref{eq:almo} is rewritten equivalently as follows:
\begin{equation} \label{eq:pakuaT}
  \left[ {\begin{array}{*{20}{c}}
  {{{\tilde J}_{1,1}}}&{{{\tilde J}_{1,2}}{W^{\mathsf{T}}}} \\ 
  {W{{\tilde J}_{1,2}}^{\mathsf{T}}}&{\left[ {\begin{array}{*{20}{c}}
  \Lambda &\vline & 0 \\ \hline 0&\vline & 0 \end{array}} \right]} 
  \end{array}} \right]\left[ {\begin{array}{*{20}{c}} {{r_a}} \\ 
  {W{r_b}} \end{array}} \right] = \left[ {\begin{array}{*{20}{c}}
  {{w_a}} \\ {W{w_b}} \end{array}} \right].
\end{equation}
Decompose $W=\sbm{W_1\\ W_2}$, where $W_1$ has $h$ rows. It is claimed 
that ${W_2}{{\tilde J}_{1,2}}^\top=0$. Indeed, the last $N-G-h$ rows 
of \eqref{eq:pakuaT} are
\begin{equation*}
    {W_2}{{\tilde J}_{1,2}}^{\mathsf{T}}{r_a} = {W_2}{w_b}.
\end{equation*}
Since $r_a$ is arbitrary, and $w_b=0$ due to \eqref{eq:loadsAA}, the
claim follows. By symmetry, it follows that the last $N-G-h$ rows and
also the last $N-G-h$ columns in the matrix on the left side of
\eqref{eq:pakuaT} are zero. Elimination of these redundant rows and
columns leads to
\begin{equation*}
\left[ {\begin{array}{*{20}{c}}
  {{{\tilde J}_{1,1}}}&{{{\tilde J}_{1,2}}{W_1}^{\mathsf{T}}} \\ 
  {{W_1}{{\tilde J}_{1,2}}^{\mathsf{T}}}&\Lambda  
\end{array}} \right]\left[ {\begin{array}{*{20}{c}}
  {{r_a}} \\ {{W_1}{r_b}} 
\end{array}} \right] = \left[ {\begin{array}{*{20}{c}}
  {{w_a}} \\ 0 \end{array}} \right].    
\end{equation*}
From the last $h$ rows it is obtained that
\begin{equation*}
   {W_1}{r_b} \m=\m  -{\Lambda^{-1}}{W_1}
   {{\tilde J}_{1,2}}^{\mathsf{T}} {r_a},    
\end{equation*}
hence
\begin{equation} \label{eq:sombulpit}
   w_a \m=\m {{\tilde J}_{1,1}}{r_a} - {{\tilde J}_{1,2}} 
   {W_1}^\top{\Lambda^{-1}}{W_1}{{\tilde J}_{1,2}}^\top r_a 
   \m=\m \bar H{r_a},
\end{equation}
where
\begin{equation*}
   \bar H \m=\m {{\tilde J}_{1,1}} - {{\tilde J}_{1,2}}{W_1}^\top
   {\Lambda^{-1}}{W_1}{{\tilde J}_{1,2}}^\top.
\end{equation*}
This matrix $\bar{H}$ is symmetric, so \eqref{eq:sombulpit} can rewritten as
\begin{multline*}
   \left[\dd P_2,\ldots \dd P_G,\dd Q_2,\ldots \dd Q_G \right]^\top 
   \m= \\ =\m \begin{bmatrix} H & \bar{H}_{1,2} \\ \bar{H}_{1,2}^\top
   & \bar{H}_{2,2} \end{bmatrix} \left[\dd\delta_2,\ldots \dd\delta_G,
   \frac{\dd V_2}{V_2^*},\ldots  \frac{\dd V_G}{V_G^*}\right]^\top.
\end{multline*}
In addition, since the generators' voltage amplitudes are constant, $\left(\dd V_2,\ldots \dd V_G\right)=0$, which leads to
\begin{equation*}
   \left[\dd P_2,\ldots \dd P_G \right]^\top \m=\m H \cdot \left[
   \dd\delta_2,\ldots \dd\delta_G \right]^\top,
\end{equation*}
with a symmetric matrix $H$. \hfill $\blacksquare$

\section{The main result} \label{sec:main_result} 

Consider the dynamic system \eqref{eq:dynamic_system1} with two or
more generators $(G\geq2)$, and a symmetric matrix of partial
derivatives $H$, as defined in \eqref{eq:matrixH}, which is calculated
at an equilibrium point~$x^*$. The primary result is the following:
\begin{itemize}
\item The equilibrium point $x^*$ is \underline{asymptotically 
   stable} if the matrix $H$ is \underline{positive definite}.
\item The equilibrium point $x^*$ is \underline{unstable} if the 
   matrix $H$ is full-rank and \underline{not positive definite}.
\end{itemize}

Theorem~\ref{main} below shows this result (implicitly), and also
provides details on the relations between the eigenvalues of the
matrices $A$ and $H$. Before stating this main theorem, recall
some definitions and terminology from linear algebra.

{\bf Notation and terminology.} Real matrices are used, but these are
often applied to complex vectors (for example, to
eigenvectors). For any square matrix $A$, $\sigma(A)$ denotes the set
of its eigenvalues. The {\em algebraic multiplicity} of an eigenvalue
$\l$ is the dimension of the space of all the eigenvectors and
generalized eigenvectors that correspond to $\l$. Let
$\langle\cdot,\cdot\rangle$ denote the {\em standard inner product} on
$\cline^n$: $\langle a,b\rangle=a_1\bar{b}_1+\ldots a_n
\overline{b}_n$. A real matrix $H$ of dimension $n\times n$ is called
{\em positive} if $H^{\mathsf{T}}=H$ and for every vector
$z\in\rline^n$, $\langle Hz,z \rangle\geq 0$. Equivalently,
$H$ is positive if $H^\top=H$ (which alone implies that
$\sigma(H)\subset\rline$) and $\sigma(H)\subset[0, \infty)$. A
positive $H$ is called {\em positive definite} if $0\not\in\sigma(H)$
(i.e., if $H$ has full rank). Equivalently, $H$ is positive definite
if $H^\top=H$ and for every nonzero vector $z\in\rline^n$,
$\langle Hz,z\rangle>0$.

Symbols $\cline_-$ and $\cline_+$ denote the open left half-plane and
the open right half-plane in $\cline$ respectively. In this paragraph,
let $A$ be a square matrix with complex entries, of dimension $n\times
n$. $A$ is called \textit{stable} if $\sigma(A)\subset\cline_-$
(equivalently, if $\lim_{t\to\infty} e^{At}=0$). The {\em strictly stable
subspace} of $A$ is defined by
\begin{equation*}
   X_-(A) \m=\m \left\{x\in\cline^n \mid \lim_{t\to\infty} 
   e^{At} x = 0 \right\}.
\end{equation*}
This space is the linear span of all the eigenvectors and generalized
eigenvectors that correspond to an eigenvalue in $\cline_-$. The
{\em strictly unstable subspace} of $A$ is $X_+(A)=X_-(-A)$. Define
the notation
\begin{equation*}
   \neg(A) \m=\m \dim X_-(A), \quad\ \pos(A) \m=\m \dim X_+(A).
\end{equation*}
From $X_-(A)\cap X_+(A)=\{0\}$ it follows that 
\begin{equation*}
   \pos(A)+\neg(A) \m\leq\m n \m.
\end{equation*}
Clearly, $\neg(X)$ is the sum of the algebraic multiplicities of the
eigenvalues of $A$ in $\cline_-$, while $\pos(A)$ is the sum of the
algebraic multiplicities of the eigenvalues of $A$ in $\cline_+$. If
$A$ is the Jacobian of a dynamic system at an equilibrium point $x^*$,
then the following facts are known (see \cite{Khali2002}): If $A$ is
stable then $x^*$ is asymptotically stable. If $A$ has an eigenvalue
in $\cline_+$, then $x^*$ is not (Lyapunov) stable.

\begin{theorem} \label{main}
Let $H$ be a symmetric matrix of dimension $(G-1)\times(G-1)$, and let
$A$ be the matrix in \eqref{eq:JacobianA}, so that its dimension is
$n\times n$, where $n=2G-1$. Then:
\begin{enumerate} 
\item For $G=1$, $A$ is stable.

For $G\geq 2$, the following facts hold: 
\item If $0\not\in\sigma(H)$, then $\pos(A) =\m \neg(H)$, $\neg(A) 
   =\m\pos(H)+G$.
\item If $A$ has an eigenvalue $\l$ with $\Re\l=0$, then $\l=0$.
\item $0\in\sigma(H)$ if and only if $0\in\sigma(A)$.
\end{enumerate}
\end{theorem}
	
\noindent {\bf Remarks concerning the theorem:}
\begin{itemize}
\item If $H>0$ then $A$ is stable. This follows from 2) by 
considering $\pos(H)=G-1$, which gives $\neg(A)=n$.
\item If $H$ is full-rank and not positive definite $(H\ngtr 0)$, 
then $A$ is unstable, and has an eigenvalue $\l\in\cline_+$. This 
follows from 2), considering the case $\neg(H)\geq 1$.
\item Concerning the stability of the equilibrium point $x^*$ for 
the nonlinear system \eqref{eq:dynamic_system1}: If $H>0$, then 
$x^*$ is (locally) asymptotically stable. If $\neg(H)>0$, then $x^*$
is not even Lyapunov stable. This follows from what was said just 
before the theorem. 
\item If $H$ is full-rank, then it is the only information needed to
determine the stability of $A$ and $x^*$.  
\item Physically, the matrix $H$ represents the
synchronizing power coefficients of the network, hence, the condition
$H > 0$ ensures that the network generates restoring torques for any
rotor angle deviation.
\end{itemize}
	
{\it Proof.} \m The proof of claim 1) is trivial: If $G=1$, then the
only state variable is $\o_1$, while $M$ and $D$ are positive numbers
and $A=-D/M$, which is clearly stable.

Next, consider claims 3) and 4), as they are required to prove claim 2). Claim 3) is proved first: Let $\l$ be an eigenvalue of $A$,
and let $q=\sbm{v\\ w}$ be the corresponding eigenvector. Then
\begin{equation} \label{eq:puki}
   U^\top Hw + \left(\l M + D\right)v \m=\m 0 \m,\qquad 
   Uv \m=\m \l w \m.
\end{equation}
In case $v=0$, this  leads  to $\l w=0$. Since by definition
$q\neq 0$, it follows that $w\neq 0$, so $\l=0$.
		
In case $v\neq 0$, multiply the first equation in \eqref{eq:puki} with
$\l$, and use the second equation to obtain
\begin{equation*}
   \l^2 Mv + \l Dv + U^{\mathsf{T}} H Uv \m=\m 0.
\end{equation*}
Take the inner product with $v$ to obtain a quadratic equation in
$\l$:
\begin{equation*}
   \l^2 \langle Mv,v \rangle + \l\langle Dv,v \rangle + 
   \langle U^{\mathsf{T}} H Uv,v \rangle \m=\m 0.
\end{equation*}
Assumed that $\l$ is purely imaginary, so
\begin{equation*}
   - |\l|^2\langle Mv,v \rangle + j|\l|\langle Dv,v \rangle + 
   \langle U^{\mathsf{T}} H Uv,v\rangle \m=\m 0.
\end{equation*}
Since $M$, $D$, and $H$ are symmetric, the imaginary part of this
equation is $|\l|\langle Dv,v\rangle=0$. But $D$ is positive
definite and $v\neq 0$, so it follows that $\l=0$.
		
Now claim 4) is addressed. First, it is shown that if $H$ has an eigenvalue 0,
then $A$ has an eigenvalue 0. To this end, let $z$ be an eigenvector
of $H$ with eigenvalue 0, such that $Hz=0$, and let
$q=\sbm{0\\ z}$. Clearly $Aq=0$, so $0\in\sigma(A)$.

Now, the other direction in claim 4) is proven, showing that if
$0\in\sigma(A)$, then $0\in\sigma(H)$. So assume that $0\in\sigma(A)$
and let $q=\sbm{v\\ w}$ be a corresponding eigenvector. Then
\begin{equation*} 
   U^\top Hw + Dv \m=\m 0 \m,\qquad Uv \m=\m 0 \m.
\end{equation*} 
Since $D$ is invertible, this leads to $w\neq 0$, and 
\begin{equation*}
   U D^{-1}U^\top Hw \m=\m 0 \m.
\end{equation*} 
Take inner product with $Hw$:
\begin{equation*}
   \langle D^{-1}U^\top Hw,U^\top H w \rangle \m=\m 0 \m.
\end{equation*} 
But $D^{-1}>0$, hence $U^\top Hw=0$. The only solution for this 
equation is $Hw=0$, so $0\in\sigma(H)$.

To prove 2) consider $G\geq 2$ and $0\not\in\sigma(H)$. It can be
verified directly that the Lyapunov equation
\begin{equation*}
   A^\top P + PA \m=\m -Q
\end{equation*} 
is satisfied by
\begin{equation} \label{Lyap}
   P \m=\m \begin{bmatrix} M  & 0 \\ 0 & H \end{bmatrix}, \quad Q 
   \m=\m \begin{bmatrix} 2D & 0 \\ 0 & 0 \end{bmatrix}.
\end{equation}
Denote $V_-=X_-(P)$ and $m=\neg(H)=\neg(P)=\dim V_-$. Similarly,
denote $V_+=X_+(P)$, so that $\dim V_+=n-m$. For any nonzero $x_0\in
V_-$, $\langle Px_0,x_0\rangle<0$. Denoting
$x(t)=e^{At}x_0$, then
\begin{equation*}
   \begin{aligned} \frac{\dd}{\dd t} \langle Px(t),x(t) \rangle 
   &=\m \langle \left[A^\top P+PA\right]x(t),x(t)\rangle 
   \m= \\ &=\m -\langle Qx(t),x(t)\rangle \leq 0 \m, \end{aligned}
\end{equation*} 
so that $\langle Px(t),x(t)\rangle\leq\langle Px_0,x_0\rangle<0$, 
which shows that $x(t)$ does not converge to 0. In other words, 
$V_-\cap X_-=\{0\}$, which implies that $\dim V_-+\dim X_-\leq n$,
so that 
\begin{equation} \label{Prighozin} 
   \dim X_-\leq n-m.
\end{equation}
Now, a similar argument is used for a nonzero $x_0\in V_+$, obtaining 
that $x(t)=e^{-At}x_0$ does not converge to zero, so that $V_+\cap 
X_+=\{0\}$. From here it is obtained that 
\begin{equation} \label{liquidated}
   \dim X_+ \leq n-(n-m) \m=\m m \m.
\end{equation}
From claim 4) it is known that $0\not\in\sigma(A)$, and hence, using claim
3), it may be concluded that $A$ has no imaginary eigenvalues. This implies that
\begin{equation*}
   \dim X_- + \dim X_+ \m=\m n.
\end{equation*}
Putting together this last equality with \eqref{Prighozin} and
\eqref{liquidated}, it follows that $\dim X_-=n-m$ and $\dim X_+=m$.
\hfill $\blacksquare$

\section{Generalization to arbitrary loads} 
\label{sec:generalize}

In this section the case when the loads are not assumed to take constant power is considered. An extension of Theorem \ref{main} to the case where the assumption that the loads in the grid consume locally constant active and reactive powers is dropped as expressed by \eqref{eq:loadsAA} is derived. The other assumptions and the notation from Sect.~\ref{sec:Problem} are kept.

The matrix $H$ can still be defined as in \eqref{eq:matrixH}, but the formula \eqref{eq:losslessAss} no longer holds. This time, $H$ will 
not be symmetric, although numerical experimentation with reasonable 
circuits shows that it will be very nearly symmetric. 

The full Jacobian matrix is introduced to describe the sensitivity of the (active) electric powers with respect to the relative rotor angles at the equilibrium point $x^*$:
\begin{equation} \label{eq:Jacobian}
   \Hscr \m=\m \frac{\partial P}{\partial \delta} 
   \bigg|_{\delta=\delta^e} \in\m \rline^{G\times (G-1)} \m,\qquad 
   \Hscr_{jk} \m=\m \frac{\partial P_j}{\partial\delta_k} \m.
\end{equation}
The last $G-1$ rows of $\Hscr$ are the matrix $H$:
$$ \Hscr \m=\m \begin{bmatrix} * \dots * \\ \scalebox{1.2}
   {$H$} \end{bmatrix}, \qquad H \in \rline^{(G-1)\times (G-1)} \m.$$

Following an argument very similar to the one leading to \eqref{eq:JacobianA}, it follows that in this new context (i.e., without assuming \eqref{eq:loadsAA}), the matrix of the linearized system is
\begin{equation} \label{Jac_with_Hscr}
   A \m=\m \left[ \begin{array}{c|c} -M^{-1}D & -M^{-1} \Hscr \\ 
   \hline U & 0_{(G-1)\times (G-1)} \end{array} \right].
\end{equation}
Note that if \eqref{eq:loadsAA} holds, then \eqref{eq:losslessAss} implies that $\Hscr=U^\top H$, and the above $A$ reduces to the one from \eqref{eq:JacobianA}. Thus, the local 
stability of the system (2) around the equilibrium $x^*$ follows
if the above $A$ is stable. 

\begin{proposition} The matrix $A$ from \eqref{Jac_with_Hscr} is 
stable if and only if $\det A\not=0$ and the nonzero solutions of the 
equation
\begin{equation} \label{Elinor}
   \det \left( Ms^2 + Ds + \Hscr U \right) \m=\m 0
\end{equation}
are in the open left half-plane $\cline_-$. 
\end{proposition}

{\it Proof.} \m Assume that $\det A\not=0$ and all the solutions of
\eqref{Elinor} are in $\cline_-$. Let $s\in\cline$ be an eigenvalue 
of $A$ and let $x=\sbm{\o\\ \delta}$ be a corresponding eigenvector,
then $x\not=0$, $s\not=0$ and
$$ -D\o -\Hscr\delta \m=\m sM\o \m,\qquad U\o \m=\m s\delta \m.$$
Multiply the first equation above with $s$ and substitute $s\delta$
from the second equation, obtaining
\begin{equation} \label{Nov_11_at_11am}
   \left[\m M s^2 + D s + \Hscr U \m\right] \o \m=\m 0 \m.
\end{equation}
Since $x\not=0$, it follows that $\omega\not=0$, so that the above equality implies \eqref{Elinor}; thus, by assumption, $s\in\mathbb{C}_-$. Hence, $A$ is stable.

Conversely, assume that $A$ is stable, then clearly $\det A\not=0$.
Let $s$ be a nonzero solution of \eqref{Elinor}. Then there exists
a vector $\o\in\cline^G$ such that $\o\not=0$ and  
\eqref{Nov_11_at_11am} holds. Denote $\delta=\frac{1}{s}U\o$, then it
is easy to see $x=\sbm{\o\\ \delta}$ is an eigenvector of $A$, 
corresponding to the eigenvalue $s$. Since $A$ is stable, it follows that
$s\in\cline_-$, as claimed. \hfill $\blacksquare$

In the following proposition sufficient conditions for 
the stability of $A$ are derived.

\begin{proposition} \label{prop:spectrum}
Assume that at least one of the following conditions holds:

$(1)$ Locally constant power loads, i.e., \eqref{eq:loadsAA}
holds.

$(2)$ $M^{-1}D=\gamma I$, where $\gamma$ is a positive number.

Then the matrix $A$ from \eqref{Jac_with_Hscr} is stable if 
\begin{equation} \label{Tareq}
   \sigma(UM^{-1}\Hscr) \m\subset\m (0,\infty) \m.
\end{equation}
\end{proposition}

{\it Proof.} \m First, assume that condition (1) holds. Then (from
\eqref{eq:losslessAss}) $\Hscr=U^\top H$ and (from Theorem
\ref{th:symm}) $H^\top=H$. Then, the condition \eqref{Tareq} becomes
\begin{equation*}
    \sigma(UM^{-1}U^{\mathsf{T}}H) \subset (0,\infty).
\end{equation*}
Notice that the above implies that Ker $H=\{0\}$, so that $\det
H\not=0$. Denoting $\bar{M}=UM^{-1}U^\top$, it follows that $\bar{M}>0$,
because $M>0$ by assumption and Ker $U^\top=\{0\}$. Denote the
nonzero spectrum of a square matrix $T$ by $\sigma_0(T)$. It is well
known from linear algebra that if $S$ and $T$ are matrices such that
$ST$ and $TS$ are both square matrices (possibly of different
dimensions), then $\sigma_0(ST)=\sigma_0(TS)$. It follows that the
condition \eqref{Tareq} is equivalent to $\sigma_0(\bar{M}^\half H
\bar{M}^\half)\subset(0,\infty)$. Since the matrices $\bar{M}^\half$ and 
$H$ have nonzero determinant, $\det\bar{M}^\half H\bar{M}^\half
\not=0$, it follows that $\sigma(\bar{M}^\half H
\bar{M}^\half)\subset(0,\infty)$. Since $\bar{M}^\half H\bar{M}
^\half$ is symmetric, this is equivalent to $H>0$. According to 
Theorem \ref{main}, $A$ is stable. 
 
Now assume that condition (2) holds. Let $\sbm{\o\\ \delta}$ be an 
eigenvector of $A$ corresponding to the eigenvalue $s$. Then
\begin{equation} \label{Jac_with_Hscr_eig}
   -M^{-1}D\o -M^{-1}\Hscr\delta \m=\m s \o \m,\qquad U \o 
   \m=\m s \delta.
\end{equation}
Multiplying the first equation by $U$ from the left and substituting
the second equation results in
\begin{equation*}
   -UM^{-1}D\o -(UM^{-1} \Hscr) \delta \m=\m s(U\o) \m=\m s^2\delta.
\end{equation*}
Assuming the special case where $M^{-1}D=\gamma I$ for some number
$\gamma>0$, the equation becomes
\begin{equation*}
   -\gamma U \o - (U M^{-1}\Hscr) \delta \m=\m s^2 \delta.
\end{equation*}
Using the second equation in \eqref{Jac_with_Hscr_eig} again, this 
may be written as
\begin{equation*}
    -\gamma (s\delta) - (U M^{-1} \Hscr) \delta \m=\m s^2 \delta.
\end{equation*}
Rearranging the terms yields
\begin{equation*}
    \left[ (UM^{-1} \Hscr) + (s^2 + s\gamma)I \right] \delta \m=\m 0.
\end{equation*}
This equation shows that for any non-trivial eigenvector with the 
corresponding eigenvalue $s$, 
$$ -(s^2 + s\gamma) \m\in\m \sigma(U M^{-1}\Hscr) \m.$$

Let $\alpha$ be any such eigenvalue of $UM^{-1}\Hscr$, which means
that $s^2+s\gamma+\alpha=0$. Assume \eqref{Tareq}, then $\alpha
>0$. Since also $\gamma>0$, the roots of $s^2+s\gamma+\alpha=0$ are in 
$\cline_-$, which means that $A$ is stable. \hfill $\blacksquare$

Numerical experiments on a variety of networks with various resistive loads 
and $\gamma$ values, confirm that the spectrum of the matrix $UM^{-1}\Hscr$ 
is real and positive for many stable systems as long as the condition 
$M^{-1}D=\gamma I$ holds. It seems that the condition $M^{-1} D = \gamma I$
implies that $\sigma(U M^{-1} \Hscr)$ is real, but there is no formal proof.

\begin{proposition} \label{rain}
Assume that the matrix $A$ from \eqref{Jac_with_Hscr} satisfies
condition $(2)$ from Proposition \ref{prop:spectrum}. If 
$UM^{-1}\Hscr$ has an eigenvalue $\alpha<0$, then $A$ has a positive
eigenvalue.
\end{proposition}

{\it Proof.} \m By assumption, $M^{-1}D=\gamma I$, with $\gamma>0$.
Let $s$ be the positive solution of the equation
\begin{equation} \label{Tareq2}
s^2 + \gamma s + \alpha \m=\m 0 \m.
\end{equation}
To prove the proposition, it suffices to show that there exists
$\sbm{\o\\ \delta}$ with $\o\in\rline^G$, $\delta\in\rline^{G-1}$,
$\delta\not=0$, such that
\begin{equation} \label{Warsaw}
   (sI-A)\bbm{\o \\ \delta} \m=\m 0 \m.
\end{equation}
Let $\delta\in\rline^{G-1}$ be such that $UM^{-1}\Hscr\delta=\alpha
\delta$, $\delta\not=0$. Since $U$ is onto, there exists $\o\in
\rline^{G-1}$ such that $U\o=s\delta$. There are many such vectors 
$\o$, since $U 1_G=0$, where $1_G$ is the vector in $\rline^G$ with 
all entries equal to 1. Thus, any value can be chosen for the
first component of $\o$, and it can be chosen such that
\begin{equation} \label{Eran_Socher}
  (s+\gamma) \o_1 + (M^{-1}\Hscr\delta)_1 \m=\m 0 \m.
\end{equation} 

Equation~\eqref{Tareq2} is now multiplied by $\delta$ and rewritten as follows:
$$ s^2 \delta + U \gamma \o + UM^{-1}\Hscr \delta \m=\m 0 \m.$$ 
Using once more that $U\o=s\delta$, results in
$$ U \left[ (s+\gamma)\o + M^{-1}\Hscr\delta \right] \m=\m 0 \m.$$ 
The $(j-1)$-th row of the above ($j\in\{2,3,\ldots G\}$) is
$$ \left[ (s+\gamma)\o + M^{-1}\Hscr\delta \right]_j -
   \left[ (s+\gamma)\o + M^{-1}\Hscr\delta \right]_1 \m=\m 0 \m.$$
Combining this with \eqref{Eran_Socher}, yields
$$ \left[ (s+\gamma)\o + M^{-1}\Hscr\delta \right]_j \m=\m 0 \m,$$
for all $j\in\{2,3,\ldots G\}$. This fact, together with 
\eqref{Eran_Socher}, means that
$$ (s+\gamma)\o + M^{-1}\Hscr\delta \m=\m 0 \m.$$
This, together with $U\o=s\delta$, means that \eqref{Warsaw} holds. 
\hfill $\blacksquare$

\section{Comparison to related previous results} 
\label{sec:compare}

Following is a comparison to some previous works that explore the 
stability of similar dynamic models of multi-machine grids (each 
generator being described by a second order swing equation) and give 
algebraic criteria for local stability. The proposed method is compared with three significant references:

* \m Machowski, Bialek and Bumby \cite{MachoBB2011} from 2012,

* \m D{\"o}rfler and Bullo \cite{DoBu:12} from 2012,

* \m Wu and Liu \cite{WuLiu} from 1986.

Consider the stability criterion given in \cite[Chapter 12]
{MachoBB2011}. The end result provided in \cite{MachoBB2011} is
similar to the one shown in the current paper: the power system
dynamic model is stable if the matrix $H$ is positive definite.
However, this result is obtained under more restrictive assumptions:
\begin{itemize}
\item The matrix $M$ of inertia constants and the matrix $D$ of
damping coefficients are assumed in \cite{MachoBB2011} to be diagonal
and normalized, such that $D_i/M_i$ is constant (independent of $i$).
This work eliminates this assumption, and it is only 
assumed that the matrices $D$ and $M$ are positive definite (not
even diagonal).
\item More importantly, careful analysis shows that in fact, the model
in \cite[Chapter 12]{MachoBB2011} corresponds to the model used in this 
work without any
load bus ($L=0$). This absence of load buses is not stated explicitly,
and is revealed only after some reasoning. Indeed, in \cite[Chapter
12]{MachoBB2011} it is assumed that the network loads are represented
by constant impedances, which are lumped within the transmission
network model, and expressed via the network admittance matrix $Y$. It
is then assumed that the transmission network, including the loads, is
lossless. This assumption is more restrictive than the
one used in the proposed framework, since here it is only assumed that the
transmission network {\bf without the loads} is lossless. This 
assumption is essential for ensuring the symmetry of the stability 
matrix, because resistive losses render the matrix $H$ non-symmetric.
Thus, the assumptions in \cite{MachoBB2011} enable only two types of 
possible loads: either the loads have no resistive components, and 
therefore consume no active power (so that they may be regarded as a 
part of the transmission network), or they are located directly on the 
generator buses, and therefore do not affect the system dynamics. Thus, 
the result in Section \ref{sec:main_result} may be regarded as a 
generalization of the one in \cite[Chapter 12]{MachoBB2011}, since locally constant power loads are allowed to appear anywhere 
within the network (in other words, arbitrary $L\geq 0$ are allowed 
instead of $L=0$). 
\end{itemize}

Still as part of the comparison with \cite{MachoBB2011}, it is noted that the proposed stability criterion is both necessary and sufficient (under the assumptions stated in Section \ref{sec:Problem}), since for a full-rank matrix $H$, an equilibrium point is stable \textit{if and only if} $H$ is positive definite. In \cite{MachoBB2011}, only the `if' part is stated. Several other properties of the eigenvalues of $H$ are also established.

Next, consider {\bf the stability result in the influential
paper \cite{DoBu:12}.} The main result concerns the well-known NRPS
model mentioned earlier, where each generator is represented by a
second order swing equation, and they are interconnected with a
linear, balanced and lossy electric network containing also the linear
loads. The derivation of the model relies on Section~3 of
\cite{WeissDL2019}. The transmission network together with the loads
is represented by its (symmetric) admittance matrix $Y$, which is
regarded as a constant complex matrix, because the frequencies are
assumed to be in a narrow range around the nominal frequency (this
assumption about $Y$ is made also in \cite{WuLiu,MachoBB2011}). The
idea in \cite{DoBu:12} is to consider the limit of this system when
the inertia constants of the generators vanish, called the nonlinear
Kuramoto (NK) model, and then show that for small enough inertia
constants and for initial states in a neighborhood of a certain
``preferred'' equilibrium point (the equilibrium point that
corresponds to the smallest relative angles), the state trajectories
of the NRPS model remain close to the state trajectories of the NK
model, which in turn is locally stable. A corrected, strengthened and
generalized version of \cite{DoBu:12} has been published in
\cite{WeissDL2019}, with a detailed explanation of what the
improvements are. For clarity, the achieved results are compared with those
in \cite{WeissDL2019}.
 
A crucial fact about the result in \cite{WeissDL2019} concerning the
NRPS model is that the inertia constants are assumed to be of the form
$M_j=\e m_j$, where $m_j>0$ is fixed and $\e>0$ is a small parameter,
and all the results hold for initial states in a neighborhood of the
preferred equilibrium, and only for sufficiently small $\e$, the upper
bound on $\e$ depending on the chosen neighborhood. Two quantities are
introduced, denoted by $\Gamma_{\min}$ and $\Gamma_{\text{crit}}$,
which depend on the parameters of the system (but not on the inertia
constants), and all the results are given under the assumption that
$\Gamma_{\min}>\Gamma_{\text{crit}}$. This assumption is rather
restrictive, according to the authors' experience. If this assumption is
satisfied, then according to part (4) of Theorem 4.1, the preferred
equilibrium point $x^*$ is exponentially attractive, i.e., all the
state trajectories in certain neighborhoods of $x^*$ converge to $x^*$
at an exponential rate, if $\e$ is sufficiently small. Compared to the achieved result, the differences are: (a) linear loads, as opposed to arbitrary loads, as discussed in Section~\ref{sec:generalize}, (b) there is no requirement for an assumption resembling $\Gamma_{\min} > \Gamma_{\text{crit}}$, (c) the inertias are not assumed to be small, (d) necessary and sufficient conditions are given (for local asymptotic stability), as opposed to sufficient conditions.

To empirically validate the theoretical differences
between the proposed criterion and the approach described in
\cite{WeissDL2019}, both methods were tested on standard IEEE
benchmark test cases with constant power loads.
Table~\ref{tab:stability_comparison} demonstrates the fundamental
limitation of the network reduction approach: while the $H$ matrix
method consistently provides definitive stability assessment across
all cases, the method described in \cite{WeissDL2019} becomes
increasingly unreliable as system complexity grows. Indeed the
sufficient condition $\Gamma_{\min}>\Gamma_{\text{crit}}$ successfully
guarantees synchronization for $\e$ small enough only for small,
tightly-coupled networks, such as ``case9'' and ``case30''. However,
for larger systems, the method in \cite{WeissDL2019} provides no
result, as observed in ``case118'', ``case1354pegase'' or
``case2869pegase''. The condition $\Gamma_{\min}>\Gamma_{\text{crit}}$
becomes overly conservative and impractical for large, sparse power
grids. In contrast, the proposed approach maintains numerical
stability and provides definitive stability assessment.

\begin{table}[htbp]
  \centering
  \caption{Comparison of stability assessment methods}
  \label{tab:stability_comparison}
  \resizebox{\columnwidth}{!}{%
  \begin{tabular}{c c c c c c} \toprule
     \textbf{test} & \textbf{buses} & \textbf{num. of} & \textbf{Proposed} 
     & \textbf{method} & $\boldsymbol{\Gamma_{\min}/
     \Gamma_{\text{crit}}}$ \\
     \textbf{case} &  & \textbf{gen/loads} & \textbf{method} & 
     \textbf{in \cite{WeissDL2019, DoBu:12}} &  \\ \midrule
        9 & 9 & 3/6 & \textbf{stable} & $\Gamma_{\min} > 
        \Gamma_{\text{crit}},$ & 4.50 \\
        &  &  &  & \textbf{stable for $\e$ small enough} &  \\
        14 & 14 & 5/9 & \textbf{stable} & $\Gamma_{\min} > 
        \Gamma_{\text{crit}},$ & 2.65 \\
        &  &  &  & \textbf{stable for $\e$ small enough} &  \\
        30 & 30 & 6/24 & \textbf{stable} & $\Gamma_{\min} > 
        \Gamma_{\text{crit}},$ & 1.18 \\
        &  &  &  & \textbf{stable for $\e$ small enough} &  \\
        118 & 118 & 54/64 & \textbf{stable} & $\Gamma_{\min} < 
        \Gamma_{\text{crit}}$ & 0.18 \\
        1354pegase & 1354 & 260/1094 & \textbf{stable} & 
        $\Gamma_{\min} < \Gamma_{\text{crit}}$ & 0.12 \\
        2869pegase & 2869 & 510/2359 & \textbf{stable} & 
        $\Gamma_{\min} < \Gamma_{\text{crit}}$ & 0.04 \\
        \bottomrule \multicolumn{6}{l}{} \\
        \end{tabular}%
} \end{table}
If $\Gamma_{\min} \leq \Gamma_{\text{crit}}$, then \cite{WeissDL2019, 
DoBu:12} give no stability result.

Finally, {\bf the results are compared with \cite{WuLiu}}, which at its
time has been a significant advance in the field, by proposing to look
at power system models using systems of differential and algebraic
equations and finding an algebraic condition for local stability.
Their model is similar to the one used in this paper, with a lossless transmission network
and with constant generator terminal voltage amplitudes, but also more
complicated: \cite{WuLiu} distinguishes between the generator internal
buses (where the synchronous internal voltage is) and the generator
terminal buses (In this work, only the internal buses are considered, while the terminal buses and the stator impedance are treated as part of the transmission network). Paper \cite{WuLiu} assumes that the loads consume active
power that is an affine function of the local frequency (equation
(34)) and reactive power that is a constant plus the reactive power
consumption of a shunt capacitor (equation (35)). They also assume
that the reactive power output at each generator internal bus is
constant. This is not intuitive, since the
generator has to maintain constant terminal voltage amplitude and a
given active power, and this seems to be one constraint too many on
each generator.

In addition, \cite{WuLiu} treats every voltage (at every node that is
not a generator terminal) as a state variable. Moreover, they treat
the phase angle at every node, except one, that is the reference, as a
state variable. This leads to more than two times more state variables
than what is needed, see equation (28). With these state
variables, the system is described by a collection of differential and
algebraic equations, (36) and (37). Under an additional mild
assumption (A7), the reference \cite{WuLiu} computes the linearization
of the aforementioned differential-algebraic system around an
equilibrium point $x$ (equations (41) and (42)) and defines a square
matrix $\tilde{\mathbf F}(x)$, which is proved to be symmetric (Lemma
1). The matrix $\tilde{\mathbf F}(x)$ is of dimension $n-1$, which is simply $2G+L-1$.  The main result of \cite{WuLiu}
(Theorem 1) has some formal resemblance with the main result of this work, namely, the
equilibrium point is (locally) stable if and only if $\tilde{\mathbf
F}(x)>0$. However, the dimension of matrix $\tilde{\mathbf F}(x)$ is
at least twice the dimension of the matrix $H$, and its definition is
more complex. In addition, there is no information in \cite{WuLiu}
about the number of unstable eigenvalues of the linearized system, in
case it is unstable.  

\vspace{-2mm}
\section{Case study 1: Loads located on the same buses as generators}
\label{sec:case_study_typeI} 

\begin{figure}[!t] \centering 
\includegraphics[width=0.5\columnwidth]{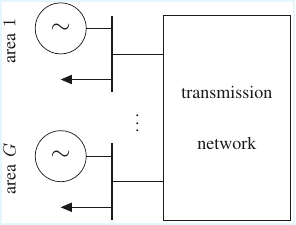}
\caption{Case study: a classic power system model, in which the loads 
are located on the same buses as the generators.}
\label{fig:type_1_net}
\vspace{-2mm}
\end{figure}

A number of case studies follow, demonstrating the usefulness of the stability criterion in Theorem~\ref{main}. The first
case study is a famous abstract power system model, in which the
system is divided into ``areas'', and each area includes one generator
and one load. This implies that the loads are located on the same
buses as the generators, see Fig.~\ref{fig:type_1_net}.

This simplified model has been studied extensively in both the power
systems and control communities for many years. In the control
community, this model is sometimes described as an instance of the
NRPS model, or analyzed in the context of a Kuramoto model
\cite{DoBu:12,DoBu:14}. In the power systems community, this model
appears in classic textbooks, for instance \cite{MachoBB2011}. Since
the mathematical properties of this model are already well known
\cite{DoBu:12,STROGATZ20001, MachoBB2011}, the main objective in this section is
not to learn something new about it, but rather to illustrate and
confirm the proposed stability criterion for this particular class of
models.

For this system, the function $f(\cdot)$ from 
\eqref{eq:dynamic_system1} can be expressed by simple formulas:
\vspace{-2mm}
\begin{equation*}
   P_i(t) \m=\m \sum_{n=1}^G {V_i}{V_n}|{y_{i,n}}|\cos(\delta_n(t)
   -\delta_i(t)+\angle y_{i,n}) + P_{L,i},
\end{equation*}
for $i=1,\ldots  G$, where $y_{i,n}$ are elements of the admittance
matrix, $V_{i}$ are the voltage amplitudes of the generators, and
$P_{L,i}$ are the active powers of the loads. For the more specific
case of a lossless transmission network that is \textit{primarily
inductive}\footnote{Lossless and primarily inductive: the transmission
network includes inductors and capacitors, and the capacitors are
small enough such that $\angle y_{i,i}=-\pi/2$, and $\angle y_{i,n} =
+\pi/2$ for $i\neq n$.}, these equations may be also written as
\vspace{-2mm}
\begin{equation} \label{Navalny}
  P_i(t) = \sum_{n=1}^G a_{i,n}\sin\left(\delta_i(t) - \delta_n(t)
  \right) + {P_{L,i}},
\end{equation}
where $a_{i,n}=a_{n,i}=V_i V_n |y_{i,n}|$. 
Computing the derivatives of these equations yields
\begin{equation*}
   h_{i,k} = \frac{\partial P_i}{\partial \delta_k} = \begin{cases}
   \text{for } i = k: \sum_{n = 1,n \ne i}^G a_{i,n}\cos\left(
   \delta_i^* - \delta_n^*\right), \\ \text{for } i \ne k: -a_{i,k}
   \cos\left( \delta_i^*-\delta_k^* \right),
\end{cases}
\end{equation*}
where $i,k\in\{2,\ldots G\}$, and the star notation marks the values
at the equilibrium point. The equality $a_{i,k}= a_{k,i}$ implies
$h_{i,k}=h_{k,i}$, so the matrix $H$ is symmetric, as expected.

According to the proposed stability criterion, for $x^*$ to be asymptotically stable, the matrix $H$ must be positive definite. To this end, consider the quadratic form $z^{\mathsf{T}}Hz$,
where $z=(z_2,\ldots z_G)^{\mathsf{T}}$ is any real vector:
\begin{equation*}
   z^{\mathsf{T}} Hz = \sum_{i=2}^G \sum_{k=2}^G h_{i,k}z_iz_k = 
   \sum_{i=2}^G \left(h_{i,i}z_i^2 + \sum\limits_{\substack{k=2, \\ 
   k\ne i }}^G h_{i,k}z_iz_k \right).
\end{equation*}
Define $b_i=a_{i,1}\cos(\delta_i^{*})$ to obtain $h_{i,i} = b_i - \sum
\limits_{\substack{k=2,\\ k\ne i }}^G h_{i,k}$, and substitute to get
\begin{equation} \label{eq:quadformABC}
   \begin{aligned} z^\top Hz &= \sum_{i=2}^G\left(b_i z_i^2-\sum
   \limits_{\substack{k=2,\\ k\ne i}}^G h_{i,k} z_i^2 + \sum
   \limits_{\substack{k=2, \\ k\ne i }}^G h_{i,k} z_{i} z_{k}\right)
   \\ &= \sum_{i=2}^G b_i z_i^2+\sum_{i=2}^G\sum\limits_{\substack
   {k=2,\\ k\ne i }}^G h_{i,k} z_i\left(z_k-z_i\right) \\ & =\sum_{i=2}
   ^G b_i z_i^2+\sum_{i=2}^G \sum_{k=i+1}^G h_{i,k} \left(2z_i z_k-
   z_i^2-z_k^2 \right) \\ & =\sum_{i=2}^G b_i z_i^2 + \sum_{i=2}^G
   \sum_{k=i+1}^G\left(-h_{i,k}\right)\left(z_i-z_k \right)^2 \\ & =
   \sum_{i=2}^G a_{i,1}\cos\left(\delta_i^*\right) z_i^2 \\ & \quad +
   \sum_{i=2}^G\sum_{k=i+1}^G a_{i,k}\cos\left( \delta_i^* - 
   \delta_k^*\right) \left(z_i-z_k\right)^{2}. \end{aligned}
\end{equation}
If this expression is $>0$ for every $z\neq 0$, then $H>0$.

\begin{figure}[!t] \centering
   \includegraphics[width=0.6\columnwidth]{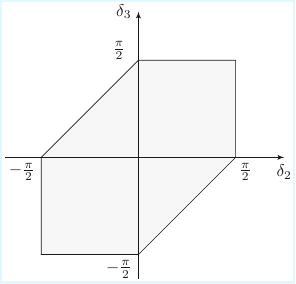}
   \caption{Case study 1, with $G=3$: If $(\delta_2,\delta_3)$ are in 
   the shaded region, then the corresponding equilibrium point is 
   asymptotically stable.} \label{fig:stability_region_G2}
\end{figure}

For $G\geq 2$, a network is called \textit{connected}
\footnote{\textbf{Connected graph}: This definition implies that every
two nodes (generators in this case) are connected to each other, directly
or indirectly, through one or more edges. In other words, there are no
``islands'' within the network. \\ \indent More formally, consider an
undirected weighted graph with adjacency matrix $A$ (see, for
instance, \cite{Bullo}), whose elements are $a_{i,k}\geq 0$ for
$i,k=1,\ldots G$, with $i\neq k$, and its main diagonal is zero. This graph is \textit{connected} if and only if
\eqref{eq:fullyAAA} holds for every real vector $z=(z_2,\ldots z_G)
^\top\ne 0$. Here is a quick informal proof for the ``if'' part of
this claim: Assume by negation that there is a group of nodes that
form an island within the network, in the sense that they are
connected to each other, but none of them is connected directly or
indirectly to the node 1. Let $B$ be the set of indices of the nodes
within this group. Define a vector $z$ as follows: $z_i=1$ if $i\in B$
and $z_i=0$ otherwise, for $i=2,\ldots G$. Clearly this vector $z$
zeros the sum in \eqref{eq:fullyAAA}, which contradicts the definition
of a connected graph. Therefore, a group of nodes that form an island
cannot exist. The ``only if'' part of the claim can be proved based on
a similar reasoning.} if for every vector $z=(z_2,\ldots z_G)^\top\neq
0$, it is true that
\begin{equation} \label{eq:fullyAAA}
  \sum\limits_{i=1}^G {\sum\limits_{k \ne i}^G {{a_{i,k}}
  {{\left( {{z_i} - {z_k}} \right)}^2}} }  \ne 0, \quad 
  \textrm{with} \quad z_1=0. 
\end{equation}
Assume that indeed $G\geq2$ and the network is connected, as above. In
this case, it can be shown that
\begin{equation} \label{eq:fullyBB}
   \sum_{i=2}^G a_{i,1} z_i^2 +\sum_{i=2}^G\sum_{k=i+1}^G a_{i,k} 
   \left(z_i-z_k\right)^{2} \m\neq\m 0,
\end{equation}
for every $z\ne 0$. 
When comparing \eqref{eq:fullyBB} to \eqref{eq:quadformABC},
it is evident that since the coefficients $a_{i,k}$ are all
non-negative, the expression in \eqref{eq:quadformABC} is strictly
positive for every $z\neq 0$, if $\cos\left(\delta_i^*\right)> 0$, 
$\cos\left(\delta_i^* - \delta_k^*\right)> 0$, for ${i,k=2,\ldots G}$. 
As a result $H$ is positive definite, and the equilibrium point 
$x^*$ is asymptotically stable, if
\begin{equation} \label{eq:anglesset}
\begin{aligned}
   -\frac{\pi}{2} &< \delta_i^*< \frac{\pi}{2}, \\
   -\frac{\pi}{2} &< \delta_i^* - \delta_k^* < \frac{\pi}{2},
\end{aligned}
\end{equation}
for $i,k=2,\ldots G$. Other stable equilibrium points can be obtained
by shifting any angle by $2\pi$. Note that \cite{WeissDL2019} and
\cite{DoBu:12} refer to the same set of angles as in
\eqref{eq:anglesset}, using the notation
$\delta\in\Delta_{\mathrm{grnd}}(\pi/2)$. The resulting region is
plotted for the case of $G=3$ in Fig.~\ref{fig:stability_region_G2}.

\section{Case study 2: Stability under the DC power-flow 
approximation} \label{sec:case_study_DC} 

The ``DC power flow'' approximate model of the power grid
is obtained when using a small angle approximation of 
\eqref{Navalny}:
\begin{equation*}
  P_i(t) = \sum_{n=1}^G a_{i,n} \left(\delta_i(t) - \delta_n(t)\right) 
  + P_{L,i},
\end{equation*}
for $i=1\ldots G$, where $a_{i,n}=a_{n,i}=V_i V_n |y_{i,n}|$, and 
$P_{L,i}$ is the equivalent load connected at bus $i$. Straight-forward differentiation yields
\begin{equation*}
  h_{i,k} = \frac{\partial P_i}{\partial \delta_k}=\begin{cases}
  \text{for } i = k: \sum_{n=1,n \ne i}^G a_{i,n}, \\
  \text{for } i \ne k: -a_{i,k}, \end{cases}
\end{equation*}
where $i,k=2,\ldots G$. Again, since $a_{i,k}= a_{k,i}$ it follows
that $h_{i,k} = h_{k,i}$, so $H$ is symmetric. Consider now the
quadratic form $z^{\mathsf{T}}Hz$, where $z=(z_2,\ldots z_G)
^{\mathsf{T}}$ is any real vector. This can be expressed as
\begin{equation} \label{eq:dc_power_flow_quad_form}
   \begin{aligned}
   z^{\mathsf{T}}Hz &= \sum_{i=2}^G \sum_{k=2}^G h_{i,k}z_iz_k  \\
   &= \sum_{i=2}^G \left(h_{i,i}z_i^2 + \sum_{k=2,k \ne i}^G h_{i,k}
   z_iz_k \right) \\
   &= \sum_{i=2}^G \left(\sum_{k=1,k \ne i}^G a_{i,k}z_i^2  - 
   \sum_{k=2,k \ne i}^G a_{i,k}z_iz_k \right) \\
   &= \sum_{i=2}^G \left(a_{i,1}z_i^2 + \sum_{k=2,k \ne i}^G a_{i,k}
   z_i\left(z_i - z_k \right)\right) \\
   &= \sum_{i=2}^G a_{i,1}z_i^2  + \sum_{i=2}^G \sum_{k=i+1}^G 
   a_{i,k}\left(z_i - z_k \right)^2. \end{aligned}
\end{equation}
If $G\geq 2$, and the transmission network is \textit{connected} (as
defined in Section~\ref{sec:case_study_typeI}), then 
the expression in \eqref{eq:dc_power_flow_quad_form} is strictly 
positive for every $z\neq 0$, which means that $H$ is positive
definite. This follows from the fact that all the coefficients
$a_{i,k}$ are non-negative. Because $H$ is positive definite, the
equilibrium point $x^{*}$ is asymptotically stable, and since the
dynamic system is linear in this example, the equilibrium point
$x^{*}$ is also globally asymptotically stable.

\section{Case study 3: Numerical results for standard ``MATPOWER'' 
case studies} \label{sec:case_study_numerical} 

In this section the proposed stability criterion is applied to several
standard case-study networks, which are taken from the software
package MatPower \cite{ZimmeMT2011} with
double-precision arithmetic. The considered case-studies are
``case9'', ``case14'', ``case30'', ''case118'', ``case1354pegase'',
and ``case2869pegase''. For these case-studies the matrix
$H$ is computed numerically, and it is shown that in all cases $H$ is symmetric and
positive-definite at the default equilibrium point $x^*$, implying
that $x^*$ is asymptotically stable. The matrix $H$ is also used to
partially explain what specific mathematical property of these
networks makes them stable. This is done by showing that most of these
networks lead to a matrix $H$ that is \textit{diagonally dominant},
see explanations below.

The computation of the $H$ matrix follows this
procedure: First, each transmission line's series resistance and bus
shunt conductance are set to zero (\texttt{BR\_R} = 0, \texttt{GS} =
0) to ensure the lossless network assumption. The equilibrium point
$x^*$ is obtained using MatPower's Newton-Raphson power flow solver
(\texttt{runpf} function) with default convergence tolerances
($10^{-8}$ for power mismatches). All generator buses are set as
$V$-$\delta$ (reference) buses to maintain constant voltage
amplitudes.

The $H$ matrix elements are computed using finite differences: each
generator angle $\delta_k$ ($k = 2,\ldots G$) is perturbed by
$\varepsilon = 10^{-3}$ radians, the power flow is re-solved, and the
partial derivatives are estimated as $h_{i,k} = \partial P_i/\partial
\delta_k \approx (P_i^u - P_i^*)/\varepsilon$, where $P_i^u$ and
$P_i^*$ are the updated and equilibrium generator powers,
respectively. This perturbation size provides sufficient numerical
accuracy while avoiding finite-precision errors. The symmetry of $H$
is verified by checking that $\|H - H^\top\|_\infty/\|H\|_\infty <
10^{-4}$. For systems with fewer than 200 generators, positive
definiteness is verified using the \texttt{eig()} function. for larger
systems, Cholesky factorization is used as a computationally efficient
alternative.

For each case-study the matrix $H$ is computed numerically, based on
the following procedure: first, the series resistance in every branch
and the shunt conductance on every bus are zeroed, to make the
transmission network lossless. At the second stage, the default
equilibrium point $x^*$ is computed by running MatPower built-in
functions. A loop is then executed in which each $\delta_k$ is slightly perturbed by the substitution $\delta_k \to \delta_k + e$, for $k=2, \ldots, G$, with a default value of $e=10^{-3}$. For each such
change the power flow is solved again, to compute the updated
generator powers $P_i^u$. The partial derivatives are then estimated
using the approximation $h_{i,k}=\partial P_i/\partial\delta_k=(P_i^u
- P_i^*)/e$. After the matrix $H$ is computed, its symmetry is verified. This numeric process is very fast, and takes a few seconds
at most for all the examined case-studies.

The equilibrium point $x^{*}$ is calculated using the Newton Raphson power flow solver provided in the MATPOWER software package. A convergence tolerance of $10^{-8}$ for power mismatches is utilized to ensure numerical precision. For the assessment of stability, the eigenvalues of the matrix $H$ are computed using the standard eig function for smaller power systems. In contrast, the Cholesky factorization is employed for larger networks such as the 2869-bus Pegase system to determine positive definiteness with high computational efficiency.

For the first case-study, ``case9'', the resulting matrix is
\begin{equation*}
   H \m=\m \begin{bmatrix} \frac{\partial P_2}{\partial\delta_2} 
   & \frac{\partial P_2}{\partial\delta_3} \\
   \frac{\partial P_3}{\partial\delta_2} & \frac{\partial P_3}
   {\partial \delta_3} \end{bmatrix} = \begin{bmatrix} \mathbf{9.31} 
   & -5.14 \\ -5.14 & \mathbf{9.45} \end{bmatrix}.
\end{equation*}
This matrix is symmetric, with eigenvalues $14.52$ and $4.24$, and
therefore it is positive-definite, which means that the default
equilibrium point $x^*$ is asymptotically stable.

Another reason that makes this matrix positive definite is that it is
\textit{strictly diagonally dominant} (SDD). Recall that a real matrix
$H$ is SDD if $\left|h_{i,i}\right| \m>\m \sum_{j\ne i} \left|h_{i,j}
\right|$. If a real symmetric matrix has this property SDD, and the 
entries on its main diagonal are positive, then this matrix is also 
positive-definite $(H>0)$, see \cite[Theorem 6.2.27]{HornJ1985}.

For ``case9'', the matrix $H$ shown above is obviously SDD, so $H>0$.
This may be qualitatively explained as follows: if for every generator
``\textit{the change in local power is dominant}'', in the sense that
$|h_{i,i}| > \sum_{j\ne i}|h_{i,j}|$, then $H$ is positive definite,
and the equilibrium point $x^*$ is stable. Intuitively, this property
will be true for many power systems, and may partially explain, based
on the proposed theorem, what makes typical power systems stable.

For the next case-study, ``case14'', a similar behaviour is observed. The
matrix $H$ is given by
\begin{equation*}
   H =\m \begin{bmatrix} \mathbf{55.04} & -13.23 & -4.13 & -1.83 \\
   -13.23 & \mathbf{17.38} & -1.81 & -1.14 \\
   -4.13 & -1.81 & \mathbf{10.43} & -2.67 \\
   -1.83 & -1.14 & -2.67 & \mathbf{6.19} \end{bmatrix}.
\end{equation*}
The eigenvalues are $59.49$, $15.02$, $10.66$ and $3.88$, so this
matrix is positive-definite, and $x^*$ is stable. As observed previously, the matrix $H$ is SDD, which explains why the equilibrium point is
stable.

\begin{figure}[htbp] \centering \label{fig:Hmat_case118} 
   \includegraphics[width=0.8\columnwidth]{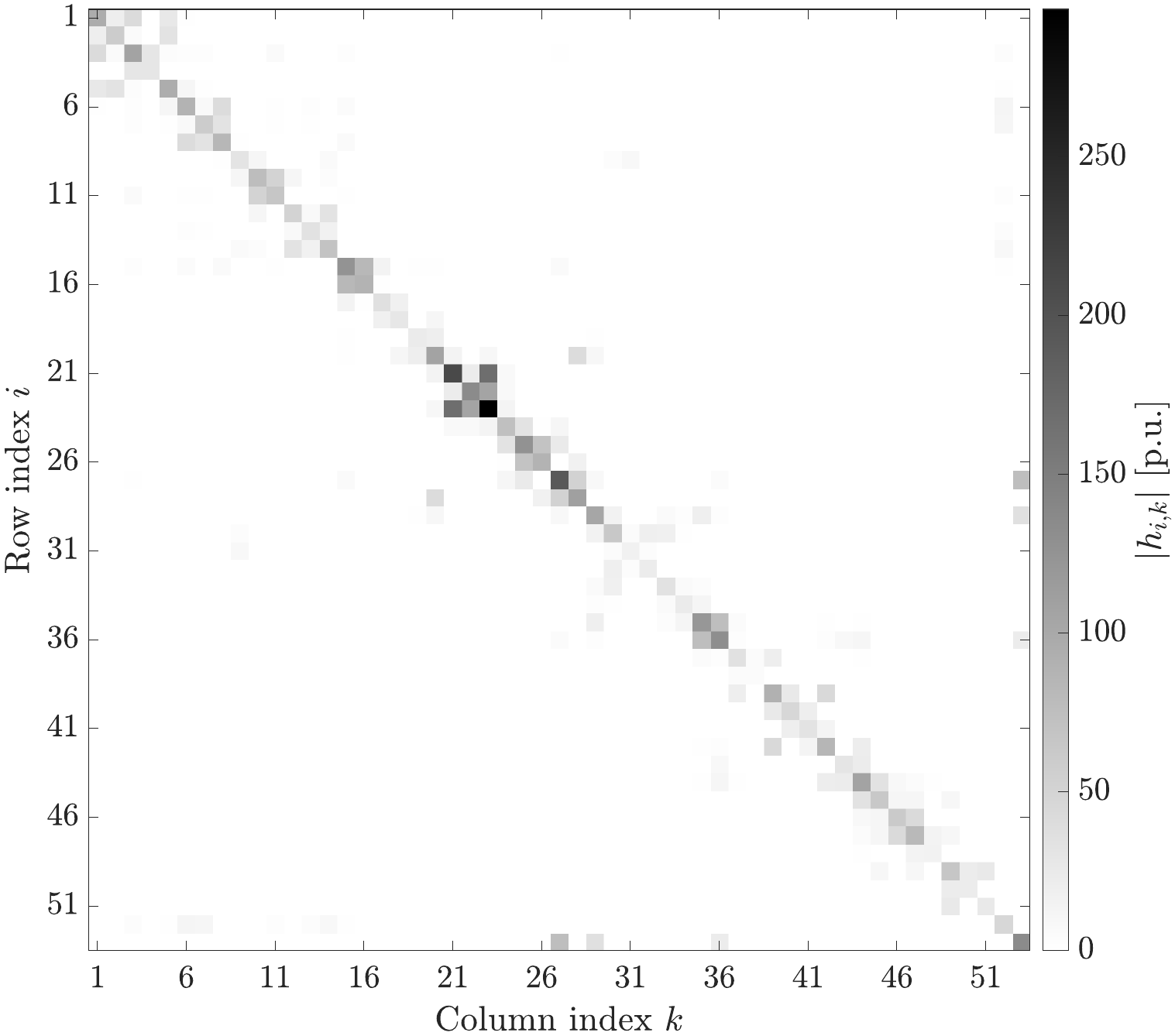}
   \caption{Schematic plot of the matrix $H$ for MatPower's 
   ``case118''. Note the dominant main diagonal.}
\end{figure}

For ``case30'', the matrix $H$ is given by
\begin{equation*}
   H =\m \begin{bmatrix} \mathbf{40.68} & -3.45 & -2.01 & -0.77 & 
   -1.62 \\ -3.45 & \mathbf{13.10} & -2.22 & -4.62 & -1.92 \\
   -2.01 & -2.22 & \mathbf{6.11} & -1.18 & -0.22 \\
   -0.77 & -4.62 & -1.18 & \mathbf{9.21} & -2.37 \\
   -1.62 & -1.92 & -0.22 & -2.37 & \mathbf{6.74} \end{bmatrix}.
\end{equation*}
The eigenvalues are $41.25$, $16.14$, $9.94$, $6.46$ and $2.06$, so
this matrix is positive-definite, and $x^*$ is stable. The matrix $H$
is again SDD.  For ``case118'' the matrix $H$ is schematically plotted
in Fig.~\ref{fig:Hmat_case118}. As shown in the figure, $H$ has a
dominant main diagonal, but in this case it is not SDD, so the
intuition provided above is not valid. Still, the eigenvalues of $H$
are all positive, so this matrix is positive-definite, and $x^*$ is
stable.

\begin{figure}[htbp] \centering 
\includegraphics[width=0.8\columnwidth]{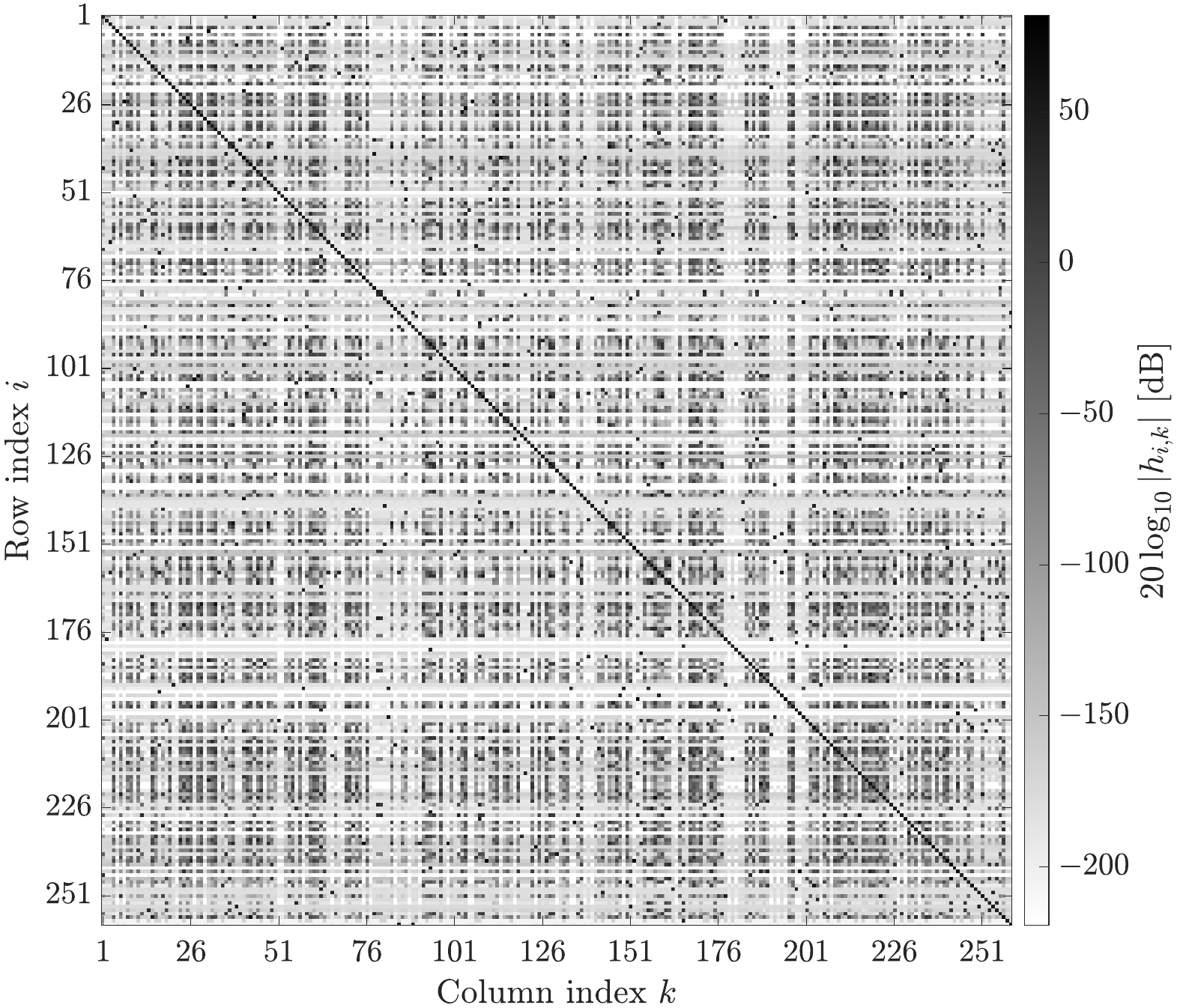}
\caption{Schematic plot of the matrix $H$, in logarithmic scale, for
MatPower's ``case1354pegase''. Note the dominant main diagonal.}
\label{fig:Hmat_case1354pegase}
\end{figure}

For ``case1354pegase'' the matrix $H$ is schematically plotted in
Fig.~\ref{fig:Hmat_case1354pegase}, in logarithmic scale. As in the
previous case, $H$ has a dominant main diagonal, but is not
SDD. Nevertheless, it can be verified using Cholesky decomposition
that this matrix is positive-definite, and $x^*$ is stable.

\begin{figure}[htbp] \centering 
   \includegraphics[width=0.8\columnwidth]{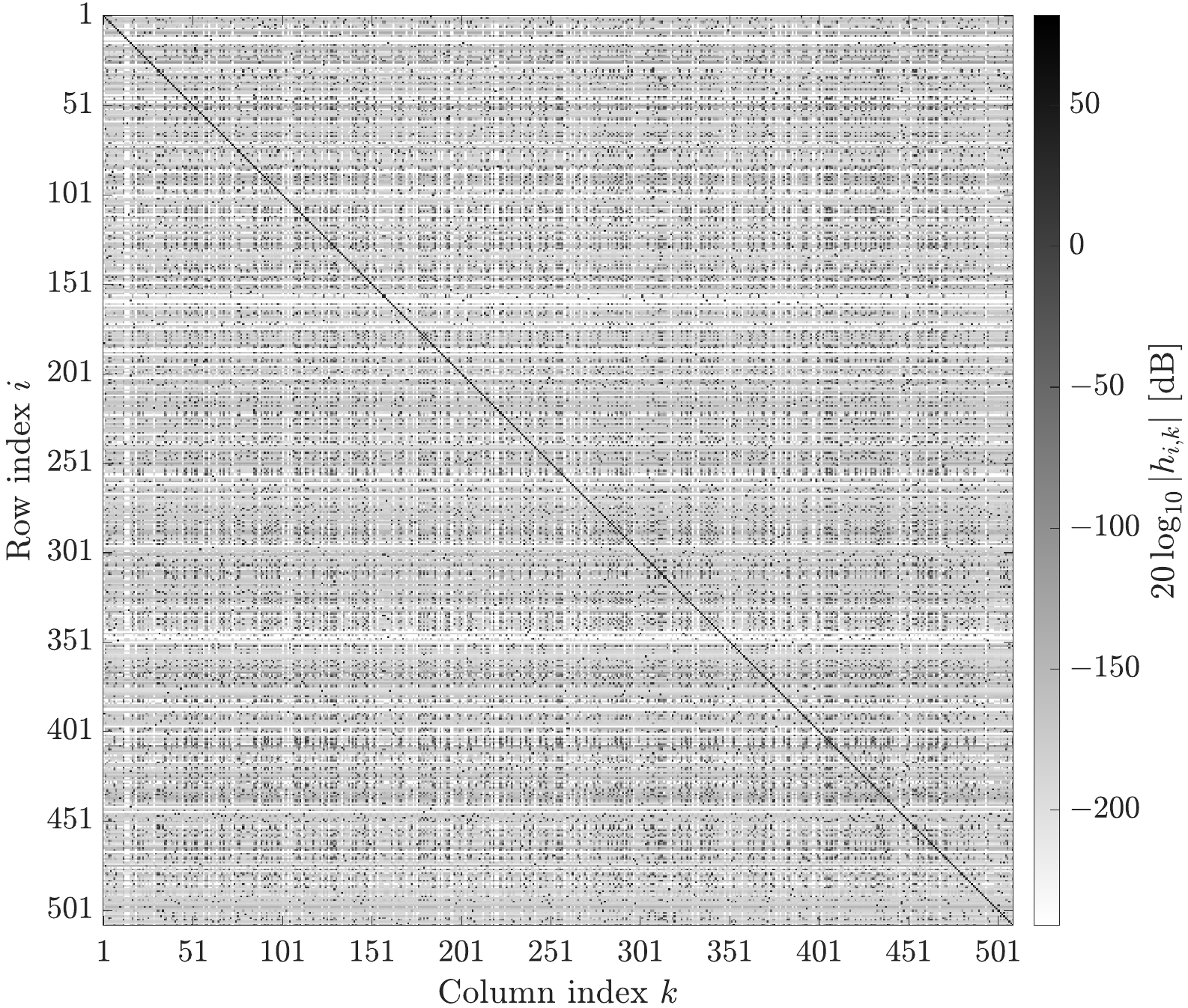}
   \caption{Schematic plot of the matrix $H$, in logarithmic scale, 
   for MatPower's ``case2869pegase''. Note the dominant main 
   diagonal.} \label{fig:Hmat_case2869pegase}
\end{figure}

Again, the ``case2869pegase'' (see Fig.~\ref{fig:Hmat_case2869pegase})
$H$ has a dominant main diagonal, but is not SDD. Nevertheless, it can
be verified using Cholesky decomposition that this matrix is
positive-definite, and $x^*$ is stable. A summary of properties of the
matrix $H$ for the above case-studies is in 
Table~\ref{tbl:H_properties}.

The system stability was examined under
scenarios typical for inverter-based generation. The
computations involved modifying the standard test cases to include
random active power injections and reactive power deviations,
examining the behavior of distributed renewable resources. In all
examined scenarios the $H>0$ and $H$ is SDD.

As detailed in Section~\ref{sec:generalize}, the stability analysis presented 
in this paper extends beyond the constant power load assumption to more general load models. When the damping and inertia matrices satisfy $M^{-1}D =
\gamma I$ with $\gamma > 0$, Proposition~\ref{prop:spectrum} provides
an alternative stability criterion through the spectrum of
$UM^{-1}\Hscr$. As noted previously, a predominant finding based on extensive numerical experiments is that the spectrum of $UM^{-1}\mathcal{H}$ is real and positive for all the stable systems that were examined.

To validate Proposition~\ref{prop:spectrum}, the same MatPower case
studies have been re-examined with modified load characteristics.
Rather than maintaining constant power consumption, the loads now
exhibit voltage-dependent behavior while $M^{-1}D=\gamma I$ is
preserved. The computational procedure remains similar to the constant
power case, with finite difference calculations of the $\Hscr$ matrix
elements following generator angle perturbations, and resistive loads
(in addition to constant power loads) are present on each load bus.

For the ``case9'' network under these modified conditions, the
eigenvalues of $UM^{-1}\Hscr$ are $12.72$ and $14.52$. Despite the $H$
matrix losing positive definiteness (it is no longer symmetric), the
system remains asymptotically stable. The Jacobian matrix $A$ exhibits
eigenvalues at $-1.0$ and complex pairs at $-0.5 \pm 3.53i$ and $-0.5
\pm 3.78i$, confirming stability.

The ``case14'' and ``case30'' networks display consistent stability
under both criteria. For ``case14'', the eigenvalues of $UM^{-1}\Hscr$
are $85.33$, $22.86$, $12.93$, and $7.31$, while $H$ is nearly
self-adjoint, with eigenvalues $59.47$, $15.02$, $10.65$, and
$3.88$. Similarly, for the ``case30'' the resulting eigenvalues of
$UM^{-1}\Hscr$ are $70.93$, $16.62$, $9.95$, $7.11$, and $6.32$. $H$
is once again nearly self-adjoint, and it has positive eigenvalues.

These numerical experiments confirm that
Proposition~\ref{prop:spectrum} encompasses a broader class of power
system models than Theorem~\ref{main}. The stability condition based
on $UM^{-1}\Hscr$ applies to systems that satisfy $M^{-1}D=\gamma I$
(which is a common assumption, as discussed in
\cite{Zargari2020Using}) regardless of whether loads satisfy the
constant power assumption.

\begin{table}[!t]
\caption{Properties of the matrix $H$ for standard case study 
   networks} \label{tbl:H_properties}
\centering
\footnotesize
\renewcommand{\arraystretch}{1.1}
\begin{tabular}{cccccc} 
\toprule
case: & 9-bus & 14-bus & 30-bus & 118-bus & 2869pegase \\
\midrule					 
$H>0$ & \checkmark & \checkmark & \checkmark & \checkmark & 
        \checkmark \\ \midrule					 
$H$ is SDD & \checkmark & \checkmark & \checkmark & X & X \\
\bottomrule
\end{tabular}
\end{table}

\section{Discussion and Conclusions} 
\label{sec:Discussion_and_Conclusions}

This paper extends existing knowledge on the small-signal stability of
time-varying phasor models of multi-machine AC power systems, offering
a generalized $H$ matrix stability criterion. For two or more
generators with a full-rank $H$, an equilibrium point $x^*$ is
asymptotically stable if $H>0$, and unstable otherwise. This result
either differs from or generalizes existing literature, as detailed in
Section~\ref{sec:compare}. Most notably, the proposed criterion
enables constant power loads to be located anywhere within the
network, even without explicit analytic power flow equations. The main
theorem is broad enough to predict stability in other classic models,
such as certain ``NRPS'' model cases (Section
\ref{sec:case_study_typeI}) and under the DC power-flow approximation
(Section \ref{sec:case_study_DC}).  Another small advantage of the
proposed theorem is that it facilitates faster stability assessments
in large-scale power systems. This is because determining if $H>0$ is
numerically simpler than computing all eigenvalues of the system
matrix $A$. This efficiency stems from two factors: first, positive
definite testing is generally more efficient in terms of run-time and
memory, with methods like the Cholesky factorization being
particularly effective for large, sparse matrices, as demonstrated in
Table~\ref{tab:matrix_performance}.  Second, the $H$ matrix's
dimension, $(G-1)\times(G-1)$, is about half of that of $A$, allowing
numerical algorithms to operate faster on $H$. Based on several case
studies conducted, the advantages of the proposed stability criterion
are demonstrated. As established in the introduction, the generalized
$H$ matrix stability criterion accurately identifies asymptotic
stability while some alternative methodologies do not yield
unequivocal results. Moreover, in contrast to approaches that rely on
the full Jacobian matrix, using the $H$ matrix, which is approximately
half the size of the system Jacobian, leads to decreased computational
time and lower memory usage.

\begin{table*}[htbp]
\centering
\caption{Computational time and memory usage of matrix $A$ and matrix 
$H$ eigenvalues calculation}
\label{tab:matrix_performance}
\resizebox{\textwidth}{!}{%
\begin{tabular}{lcccccc}
\toprule
& \multicolumn{3}{c}{\textbf{Matrix $A$}} & 
  \multicolumn{3}{c}{\textbf{Matrix $H$}} \\
\cmidrule(lr){2-4} \cmidrule(lr){5-7}
\textbf{Case} & \textbf{Dimensions} & \textbf{Comp. Time (s)} & 
  \textbf{Memory (bytes)} & \textbf{Dimensions} & \textbf{Comp. Time 
  (s)} & \textbf{Memory (bytes)} \\
\midrule
Case9 & $5 \times 5$ & 0.0001 & 200 & $2 \times 2$ & 0.0001 & 32 \\
Case14 & $9 \times 9$ & 0.0001 & 648 & $4 \times 4$ & 0.0001 & 128 \\
Case30 & $11 \times 11$ & 0.0001 & 968 & $5 \times 5$ & 0.0001 & 200 \\
Case118 & $107 \times 107$ & 0.0027 & 91,592 & $53 \times 53$ & 0.0001 
  & 22,472 \\
Case2869pegase & $1019 \times 1019$ & 0.3756 & 8,306,888 & $509 \times 
  509$ & 0.0011 & 2,072,648 \\
Case1354pegase & $519 \times 519$ & 0.0846 & 2,154,888 & $259 \times 
  259$ & 0.0004 & 536,648 \\
\bottomrule
\end{tabular}%
}
\end{table*}

\bibliographystyle{ieeetr}
\bibliography{publicationsAbrv, refs}

\vspace{-15mm}
\begin{IEEEbiography}[{\includegraphics[width=1in,height=1.2in,clip,
keepaspectratio]{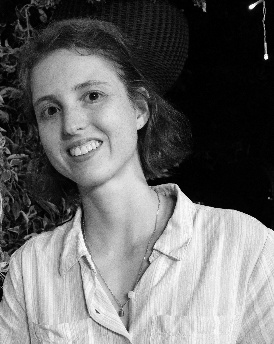}}]{Elinor Ginzburg-Ganz} received
her BSc from the Faculty of Electrical and Computer Engineering at the
Technion in 2023. She is currently pursuing her PhD on a direct track
in the Faculty of Electrical and Computer Engineering, also at the
Technion. Her research interests include energy and power systems,
optimal control, and machine learning. Ms. Ginzburg Ganz was the
recipient of The Vatat Student Research Award in the Field of
Data-Sciences in 2024, and The Freud Award for Excellence in Research
in 2025.
\end{IEEEbiography}

\vspace{-15mm}
\begin{IEEEbiography}[{\includegraphics[width=1in,height=1.2in,clip,
keepaspectratio]{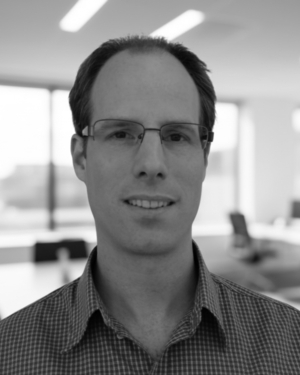}}]{Yoash Levron} (SM'18) is an
associate professor at the Technion---Israel Institute of Technology
and a faculty member since 2014. His research interests are power
systems, focusing especially on dynamics, control and artificial
intelligence, and power electronics. He has published over 100 papers
in professional journals and won several awards, among them the Masao
Horiba award, best-of-the-best paper award in the IEEE PES General
Meeting Conference, the best paper award (second place) in the IEEE
Transactions on Power Electronic, the Uzi and Michal Halevy Award for
Innovative Applied Engineering, the Technion Commendation for
Excellence in Teaching, and the Israel Defence Prize.
\end{IEEEbiography}

\vspace{-15mm}
\begin{IEEEbiography}[{\includegraphics[width=1in,height=1.25in,clip,
keepaspectratio]{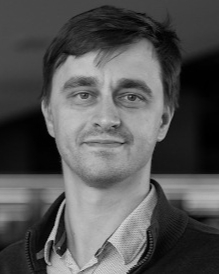}}]{Juri Belikov} (SM'19) received his
B.S. degree in mathematics from Tallinn University, and his M.S. and
Ph.D. degrees in computer and systems engineering from Tallinn
University of Technology in 2006, 2008, and 2012, respectively. In the
years 2015-2017 he was a Postdoctoral Fellow in the Faculty of
Mechanical Engineering and Andrew \& Erna Viterbi Faculty of
Electrical Engineering, Technion---Israel Institute of Technology. He
is currently an associate professor in the Department of Software
Science, Tallinn University of Technology. His main research interests
lie on the edge of nonlinear control theory, power systems, and
computer science.
\end{IEEEbiography}

\vspace{-15mm}
\begin{IEEEbiography}[{\includegraphics[width=1in,height=1.25in,clip,
keepaspectratio]{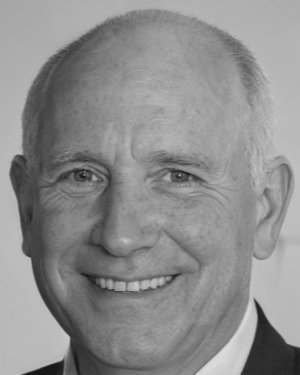}}]{George Weiss} received the MEng
degree in control engineering from the Polytechnic Institute of
Bucharest, Romania, in 1981, and the PhD degree in applied mathematics
from Weizmann Institute, Rehovot, Israel, in 1989. He has worked in
many places, including the University of Exeter, UK, and Imperial
College London, UK. His research interests include distributed
parameter systems, operator semigroups, power electronics, and the
transient stability of the power grid.
\end{IEEEbiography}

\EOD
\end{document}